\documentclass{JHEP3}

  \usepackage{latexsym,bm,amsmath,amssymb,amsfonts}
  \usepackage{epsfig,graphics,graphicx}
  \usepackage{slashed}

  \long\def\comment#1{ }

  \newcommand{\abar}{\bar{\alpha}_s}
  
  \newcommand{\del}{\partial}

  \newcommand{\mcal}{\mathcal}
  \newcommand{\rme}{{\rm e}}
  \newcommand{\rmd}{{\rm d}}   

  \newcommand{\Tr}{{\rm Tr}}

  \newcommand{\Lam}{\Lambda_{{\rm QCD}}}
  
  \newcommand{\nn}{\nonumber\\}
   
  \newcommand{\order}[1]{\mcal{O}{(#1)}}

  \newcommand{\beq}{\begin{eqnarray}}
  \newcommand{\eeq}{\end{eqnarray}}
  
 \def\simge{\mathrel{%
   \rlap{\raise 0.511ex \hbox{$>$}}{\lower 0.511ex \hbox{$\sim$}}}}
\def\simle{\mathrel{
   \rlap{\raise 0.511ex \hbox{$<$}}{\lower 0.511ex \hbox{$\sim$}}}}

\vspace{1.5cm}

\title{\rm \LARGE Deep inelastic scattering at strong coupling
from gauge/string duality : the saturation line}

\author{Y.~Hatta and E. Iancu\\Service de Physique Th\'eorique, CEA Saclay,
  F-91191 Gif-sur-Yvette, France\\
        E-mail: \email
{Yoshitaka.Hatta@cea.fr}, \email{Edmond.Iancu@cea.fr}}
\author{A. H.~Mueller\\Department of Physics, Columbia University, New York, NY
10027, USA\\
        E-mail: \email{amh@phys.columbia.edu}}

\abstract{For gauge theories which admit a dual string description, we analyze deep
inelastic scattering at strong 't~Hooft coupling and high energy, in the
vicinity of the unitarity limit. We discuss the onset of unitarity
corrections and determine the saturation line which separates weak
scattering from strong scattering in the parameter space of rapidity and
photon virtuality. We discover that the approach towards unitarity
proceeds through two different mechanisms, depending upon the photon
virtuality $Q^2$: single Pomeron exchange at relatively low $Q^2$ and,
respectively, multiple graviton exchanges at higher $Q^2$. This implies
that the total cross--section at high energy and large $Q^2$ is dominated
by diffractive processes. This is furthermore suggestive of a partonic
description where all the partons have transverse momenta below the
saturation momentum and occupation numbers of order one.
}

\begin{document}

\section{Introduction}
\setcounter{equation}{0} \label{SECT_INTRO}

High energy, small angle, scattering has always been the subject of much
attention, since the early days of elementary particle physics. While the
phenomenological Regge theory, centered around the `soft Pomeron', has
met with considerable success in reproducing the experimental data for
hadronic collisions, a fundamental understanding of this theory rooted in
QCD is still lacking and seems to even transcend our present capabilities
--- as neither perturbative QCD, nor lattice gauge theory, can be applied
in the relevant `Regge kinematics' (high energy, but soft momentum
transfer).

In perturbative QCD, the resummation of the logarithmically enhanced
diagrams leads to the BFKL, or `hard', Pomeron \cite{BFKL} which predicts
a power--like growth with $s$ for the gluon occupation number and the
scattering amplitudes. The growth is however faster than observed in
experiments, and is also problematic from a conceptual point of view as
it violates unitarity. Within the past decade there has been considerable
activity and some important progress towards understanding the
unitarization of the BFKL Pomeron within pQCD. These efforts have been
mainly oriented in two directions: \texttt{(i)} the calculation
\cite{camici} and proper implementation \cite{cia} of the
next--to--leading order corrections to the linear BFKL equation, and
\texttt{(ii)} the inclusion of the non--linear effects, like multiple
scattering and gluon saturation, responsible for the unitarization of the
scattering amplitudes \cite{B,K,AM99,JKLW,CGC,SAT}.

Whereas weak coupling techniques are undoubtedly useful for the
high--energy problems which involve, at least, one hard momentum scale
--- so like deep inelastic scattering at small $x$ and large photon
virtuality $Q^2$ ($s\gg Q^2\gg\Lam^2$), or hadronic scattering with large
momentum transfer ($s\gg |t|\gg\Lam^2$) ---, on the other hand it remains
unclear whether such approaches can shed any light on the `soft'
($|t|\simle \Lam^2$) aspects of Regge phenomenology. For instance, is
there any relation between the soft and the hard Pomerons, and if so,
then how to describe the transition between the two ? Such questions are
particularly difficult because of our inability to explore the transitive
region where the QCD coupling constant becomes strong. It is therefore
interesting to study model field theories which are analytically
accessible at strong coupling and which share some of the basic features
of QCD at high energy --- like the existence of a partonic description at
weak coupling and a Regge--like behavior at high energy, for any value of
the coupling.

${\mathcal N}=4$ supersymmetric Yang--Mills (SYM) theory with `color'
gauge group SU($N_c$) and the `t Hooft coupling constant $\lambda=g^2N_c$
treated as a free parameter offers a particularly interesting theoretical
laboratory for this purpose. At weak coupling, the high energy behavior
is just like in perturbative QCD --- in particular, to lowest order, the
BFKL equation is formally identical in QCD and ${\mathcal N}=4$ SYM
\cite{lip} --- while at strong coupling and in the large $N_c$ limit the
AdS/CFT correspondence \cite{malda} maps the theory onto a weakly coupled
superstring theory which naturally exhibits Regge behavior (see, e.g.,
\cite{amati,anna2,hard,dis,pomeron}). Such a change of behavior with
increasing the coupling strength bears some resemblance to the transition
from the hard Pomeron to the soft Pomeron in QCD, although some important
differences persists. (For instance, when moving from weak to strong
coupling, the Pomeron intercept decreases in QCD, while it increases in
the ${\mathcal N}=4$ SYM theory.) Thus one can hope that some useful
insights for the corresponding QCD problems may be gleaned from a study
of the Regge regime in ${\mathcal N}=4$ SYM theory and its variants.

In this paper we study deep inelastic scattering (DIS) in strongly
coupled ${\mathcal N}=4$ theory at small--$x$ where $x\simeq Q^2/s$ is
the usual Bjorken variable. The basic formulation was given by Polchinski
and Strassler \cite{dis} in the strict large--$N_c$ limit ($N_c\to
\infty$ at fixed energy), where however no unitarity issue arises: being
suppressed by a factor $1/N_c^2$, the elementary scattering amplitude is
always smaller than one, and is dominated by the exchange of a single
`Pomeron'. In this strong--coupling context, the `Pomeron' is the
$t$--channel object exchanged at tree--level and at high energy in the
dual string theory \cite{pomeron} --- a reggeized graviton propagating in
the curved space--time that the string theory lives in and which is
asymptotically $AdS_5\times S^5$.

Thus, in order to study unitarization, one should rather consider the
high--energy limit $s\to\infty$ at large, but fixed, values of $N_c$, and
this is what we shall do in this paper. This is again similar to QCD (at
large $N_c$), where the unitarity corrections are suppressed by inverse
powers of $N_c$, hence a flexible definition of the `large--$N_c$ limit'
is necessary to observe such phenomena. The problem of unitarization in
the context of string theory is notoriously difficult, even at weak
(string) coupling $g_s^2\propto 1/N_c^2$, because of the need to resum
multi--loop string amplitudes to all orders, and here we shall not
attempt to provide a full analysis in that sense (see however the
approaches in Refs. \cite{amati,cor,tan}). Rather, our goal is more
modest and also more pragmatic: Following again the example of
weakly--coupled QCD --- where it has been shown
\cite{GLR,SCALING,MT02,MP03} that one can follow the onset of unitarity
corrections via an appropriate extrapolation of the results at weak
scattering (in that case, the single BFKL--Pomeron exchange) ---, we
shall compute, at strong coupling, the {\em saturation line} which
separates the weak--scattering from the strong--scattering regions in the
kinematical plane $(x,Q^2)$ for DIS.

Specifically, while staying in the weak--scattering regime --- which at
strong coupling, like at weak coupling, corresponds to sufficiently large
values of $Q^2$ for a given value $x\ll 1$ ---, we shall first identify
those contributions to the scattering amplitude which become dominant
when extrapolating towards the unitarity limit, and then deduce the
saturation line as the curve $Q^2_s(x)$ along which the amplitude is
constant and of $\order{1}$. What at a first sight may look as a rather
straightforward exercise in unitarizing the single--Pomeron exchange of
Refs. \cite{dis,pomeron}, turns out to be quite subtle, because of
several issues:

First, the formulation of the unitarity constraint is rather non--trivial
in the context of DIS, where the scattering is initiated by an
off--shell, electromagnetic, current (actually, an ${\mathcal R}$-current
in ${\mathcal N}=4$ SYM). Within the context of pQCD, this issue is most
conveniently solved by using the dipole factorization of DIS at small $x$
\cite{NZ91,DIPOLE}, where one can unambiguously identify the on--shell
hadronic state which scatters off the proton, and for which the unitarity
bound can be properly formulated: the quark--antiquark excitation of the
virtual photon, or `color dipole' (see also Sect. 2 below, and Refs.
\cite{FR97,SATreviews} for more detailed discussions). It turns out that
a formally similar factorization holds at strong coupling as well, with
the color dipole replaced by a vector--field fluctuation of the
background metric (the `${\mathcal R}$-boson'), which propagates
on--shell before eventually scattering off the `dilaton' target
\cite{dis}. This makes it natural to enforce the unitarity constraint at
the level of the elementary scattering between the ${\mathcal R}$-boson
and the dilaton. This constraint will determine our saturation line.

Second, unlike what happens in pQCD, where the single (BFKL) Pomeron
exchange controls the dipole amplitude in the vicinity of the unitarity
limit for all the relevant values of $x$ and $Q^2$, the situation at
strong coupling appears to be more complex, and also more interesting:
there, the single--Pomeron exchange represents the dominant contribution
(when approaching the saturation line, once again) only for sufficiently
low values of $Q^2$, up to a critical value $Q^2_c\propto \exp\{4\ln
N_c^2/\sqrt{\lambda}\}$. But for $Q^2\! > \! Q^2_c$, a new mechanism
takes up the leading role, namely the {\em diffractive} scattering via
the exchange of two, or more, elementary (massless) gravitons. Note that
$Q^2_c\to\infty$ in the strict large--$N_c$ limit, which explains why
this new mechanism has not been included in the analysis in Ref.
\cite{dis}. Interestingly, the picture of DIS at high energy and $Q^2\! >
\! Q^2_c$ that we shall arrive is quite close to that of Regge scattering
in superstring theory in flat space, as studied by Amati, Ciafaloni and
Veneziano \cite{amati}.

In fact, for a strongly--coupled gauge theory having a dual string
description, the graviton exchange is naturally more effective, at high
energy and large $Q^2$, than the single--Pomeron one: the amplitude for
one--graviton exchange grows very fast with the energy, as $s\sim 1/x$,
because the graviton has spin $j=2$, and, moreover, its long--range
non--locality can efficiently match the large separation in scales
between the highly virtual ${\mathcal R}$-current and the dilaton. (Of
course, being real, the single graviton exchange cannot contribute to the
DIS structure functions, but multiple exchanges can do so, via
diffractive final states where the outgoing fields are separated in
rapidity.) The reason why the single--Pomeron exchange is nevertheless
found to dominate at relatively low $Q^2$ is because, in the $AdS_5$
geometry, the massless graviton is turned off at very short separations
(in the radial direction of $AdS_5$), by the curvature of space--time.
Or, by the holographic principle of AdS/CFT (which becomes manifest in
the factorization of DIS alluded to above), small separations in $AdS_5$
correspond to low virtualities $Q^2$ on the gauge theory side.

This transition, from a single--Pomeron to multiple graviton exchanges,
with increasing $Q^2$ has important consequences on the shape of the
saturation line, correspondingly obtained as the juxtaposition of two
curves which smoothly match with each other at $Q^2 = Q^2_c$, and also on
the behaviour of the DIS cross--section in the vicinity of this line. We
thus find that, at large $Q^2\! > \! Q^2_c$, the structure function
$F_2(x,Q^2)$ is strongly peaked at the saturation momentum --- namely,
$F_2\propto N_c^2Q^2$ when $Q^2 \! < \! Q^2_s(x)$, but $F_2\propto
1/(N_c^2Q^2)$ when $Q^2 \! \gg \! Q^2_s(x)$ ---, which is suggestive of a
partonic interpretation with all the partons living at transverse momenta
$k_\perp\simle Q_s(x)$ and having occupation numbers of $\order{1}$. This
is further supported by an analysis of the `sum--rule' $\int \rmd x
F_2(x,Q^2)$ which expresses the energy--momentum conservation: at large
$Q^2$, this sum--rule appears to be saturated by values of $x$ near the
saturation line, i.e., the largest values of $x$ at which one can still
find the partons. Needless to say, at strong coupling such a partonic
language must be used with care, since the standard relations between
structure functions and parton distributions hold, strictly speaking,
only in perturbation theory. It is nevertheless interesting that the
picture of parton saturation suggested by our analysis appears as a
natural continuation towards strong coupling of the corresponding results
as weak coupling, as obtained in pQCD
\cite{GLR,MQ85,MV,B,K,AM99,JKLW,CGC,SAT,SATreviews}.

This paper is organized as follows: In Section~2, we give a
self--contained review of high--energy DIS in QCD at weak coupling, in
the framework of dipole factorization. The focus is on the definition and
the construction of the saturation line, with the purpose of facilitating
the subsequent discussion at strong coupling, and also the comparison
between the weak and, respectively, strong--coupling scenarios. In
Section~3, we describe our setup of DIS at strong coupling following
\cite{dis} and demonstrate a limitation of the single--Pomeron exchange
approximation, which fails to saturate the energy--momentum sum rule at
large $Q^2$. This points out towards the importance of the massless
graviton exchanges, which are then discussed at length in Section~4. On
this occasion, we use and extend the `Pomeron' approach initiated in
Refs. \cite{hard,dis} and more systematically developed in
\cite{pomeron}, with the purpose of demonstrating the reemergence of the
massless graviton propagator at large $Q^2$. We then describe multiple
graviton exchanges at a heuristic level, using in particular the results
in Refs. \cite{amati,cor,tan} to deduce the onset of unitarity
corrections. We summarize our results for the saturation line in a `phase
diagram', Fig.~\ref{phase2}, for gauge theories at high energy and strong
't Hooft coupling, to be compared with the corresponding diagram at weak
coupling in Fig.~\ref{FigQCD}. Section~5 is devoted to a discussion of
our results, in relation with a possible partonic interpretation.

\section{Saturation momentum and geometric scaling
in perturbative QCD} \setcounter{equation}{0} \label{SECT_BFKL}

As mentioned in the Introduction, deep inelastic lepton--hadron
scattering (DIS) at relatively high virtuality $Q^2\gg \Lam^2$ for the
exchanged photon and in the high--energy limit $s\gg Q^2$ represents a
rather clean laboratory for a theoretical study of unitarization in the
framework of perturbative QCD --- in particular, for the calculation of
the saturation momentum. More precisely, in what follows we shall work in
the `small--$x$' regime of DIS, characterized by
   \beq\label{QCD_HE}
   \alpha_s\ll 1,\qquad x\,\equiv\,\frac{Q^2}{s}\ll 1,\qquad\mbox{and}
 \qquad\alpha_s\,\ln\frac{1}{x}\,\ge\,1  \,.
 \eeq
A `lowest--order calculation' in this regime requires an all--order
resummation of the perturbative contributions of order
$(\alpha_s\ln(1/x))^n$, with $n\ge 1$. This resummation can be performed
by solving appropriate evolution equations, constructed within pQCD,
which are generally non--linear, with the non--linear terms accounting
for the `unitarity corrections' --- gluon saturation and multiple
scattering \cite{B,K,JKLW,CGC,PLOOP}. However, a linear approximation ---
the BFKL equation \cite{BFKL,DIPOLE,camici} --- holds within an
intermediate range of energies, in which the non--linear effects remain
parametrically small. The boundary of the validity region for this
approximation in the kinematical plane $(x,Q^2)$ is known as the
`saturation line'; this is the curve $Q^2=Q^2_s(x)$ along which the
scattering amplitude is constant and of order one. Remarkably, it turns
out that the position of this line can be determined via calculations
based on the BFKL approximation alone, without a detailed knowledge of
the non--linear dynamics responsible for unitarization
\cite{SCALING,MT02,MP03}. In what follows, we shall briefly review this
calculation to `leading--logarithmic accuracy' (LLA), i.e., at the level
of the leading--order BFKL equation. (See also Ref. \cite{DT02,BKNLO} for
more accurate analyses, including NLO effects.)

To LLA, the cross--sections in QCD at high energy can be computed within
a factorization scheme known as `$k_T$--factorization' (see
\cite{collins,FR97}), which is consistent with the non--locality of the
high--energy evolution in transverse coordinates. When applied to DIS,
this is most suggestively written as {\em dipole factorization} : the
virtual photon $\gamma^*$ fluctuates into a quark--antiquark pair in a
color--singlet state, or `color dipole', which then scatters off the
gluon field inside the target. More precisely, the DIS structure function
$F_2$ is computed as
 \beq
  F_2(x,Q^2)\,&=&\, \frac{Q^2}{4\pi^2
 \alpha_{\rm em}} \,\sigma_{\gamma^*p}(Q^2,x), \label{fff2}
 \eeq
with $\sigma_{\gamma^*p}$ the total $\gamma^*p$ cross--section :
 \beq\label{sigmaLT}
 \sigma_{\gamma^*p}(Q^2,x)\,&=&\,\int_0^1 \rmd z \int \rmd^2 {\bm r}
 \sum_{a=T,L}\vert \Psi_{a}(z,r)\vert^2 \, \sigma_{\rm
 dipole}(r,x), \eeq
where the subscript $a$ refers to the polarization of the virtual photon
(transverse or longitudinal), and $\Psi_{a}(z,r)$ is the wavefunction
describing the dissociation of the virtual photon into a $q\bar q$ pair
with transverse size $r=|{\bm r}|$ and where the quark (antiquark) takes
away a fraction $z$ (respectively, $1-z$) of the photon energy. For
massless quarks, one finds \cite{NZ91}
 \beq \label{PsiTL} \vert \Psi_{T}(z,r)\vert^2
 &=\,& \mathcal{F}\,[z^2 +(1-z)^2] \bar Q^2  K_1^2(\bar Q r),\nn
 \vert \Psi_{L}(z,r)\vert^2 &=\,& 4\mathcal{F}\,
                      Q^2 z^2 (1-z)^2 K_0^2(\bar Q r)\,, \eeq
where $\bar Q^2 \equiv z (1-z)Q^2$, $K_0$ and $K_1$ are modified Bessel
functions, and $\mathcal{F}\equiv (N_c \alpha_{\rm em}/2 \pi^2) \sum_ f
e_f^2$, with $e_f$ the electric charge of the quark with flavor $f$. The
factor of $N_c$ comes from the sum over the color degrees of freedom of
the quarks. The integration over $r$ in Eq.~(\ref{sigmaLT}) is
effectively restricted to $r < 1/\bar Q$ by the Bessel functions, which
are exponentially suppressed for $\bar Q r\gg 1$.

Furthermore, Eq.~(\ref{sigmaLT}) involves the total cross--section
$\sigma_{\rm dipole}(x,r)$ for the scattering between a color dipole with
transverse size $r$ and the hadronic target. For the present purposes, it
is convenient to take the target itself as a collection of dipoles,
succinctly referred to as the {\em onium}. In its own rest frame, the
onium is a collection of dipoles with sizes $r'\le R_0$ distributed
according to a density function $n(r')$ and which are homogeneously
distributed in impact parameter space within a disk of radius $R_0$. Note
that $R_0$ plays the role of an `infrared cutoff' for this dipole
distribution, in the sense that this is the maximally allowed dipole
size. To ensure applicability of perturbation theory, we shall assume
that $1/R_0^2 \gg \Lam^2$ (but this assumption is often relaxed in
phenomenological studies of DIS, in which case $R_0$ is identified with
the proton radius). Also, it is convenient to choose $Q^2\gg 1/R_0^2$, in
which case the integral in Eq.~(\ref{sigmaLT}) is dominated by relatively
small dipole sizes $r$ (for the projectile dipole), such that $r\ll R_0$.

To the accuracy of interest, the dipole--onium cross--section can be
evaluated as
 \beq\label{sigmaDIP}
 \sigma_{\rm dipole}(r,\tau)\,=\,\int_0^\infty\frac{\rmd r'^2}{r'^2}\,
 \sigma_{\rm DD}(r,r',\tau)\,n(r')\,,\eeq
where $\tau\equiv \ln(1/x)$ is the `rapidity' and $\sigma_{\rm
DD}(r,r',\tau)$ is the total cross--section for the scattering between
two dipoles with transverse sizes $r$ and $r'$ separated by a rapidity
interval $\tau$. The dipole density is chosen as
  \beq\label{nr}
 n(r')\,=\,
 \left(\frac{r^{\prime 2}}{R_0^2}\right)^{\Delta} \Theta(R_0-r'),\eeq
where the power $\Delta$ plays the role of an `anomalous dimension':
indeed, the standard situation in perturbative QCD (at least for
sufficiently small $r'$) is $\Delta=0$, corresponding to the
bremsstrahlung of small dipoles. The associated distribution is not
normalizable (it exhibits a logarithmic divergence at $r'\to 0$), but
this poses no problem for the calculation of the cross--section
(\ref{sigmaDIP}) since $\sigma_{\rm DD}(r,r',\tau)$ vanishes sufficiently
fast when $r'\to 0$ (see below). A normalizable hadronic state would have
$\Delta>0$, in which case $n(r')$, once properly normalized, would have
the interpretation of the probability density to find a dipole of size
$r'$ in the target. In what follows, we shall consider the cases where
$\Delta$ is positive or zero.

We shall soon compute the cross--section (\ref{sigmaDIP}) in the BFKL
approximation and then use the result, in association with the unitarity
constraint, to deduce an equation for the saturation line. In preparation
of this, it is useful to open a parenthesis and explain our specific use
of the unitarity constraint for the present purposes. (A similar use will
be made in the subsequent sections in the context of DIS at strong
coupling.)

The unitarity bound is most easily formulated in terms of the scattering
amplitude $T(r,b, \tau)$ at fixed impact parameter $b$. It then reads
$T(r,b, \tau)\le 1$, where the upper bound $T=1$ (the `black disk limit')
describes a situation where the scattering occurs with probability one.
Note that we use conventions in which the $S$--matrix is written as
$S=1-T$, with $T$ a real quantity at high energy. (That is, our $T$
corresponds to the imaginary part of the conventionally--defined
`scattering amplitude', which becomes predominantly imaginary at high
energy.) However, the previous formul\ae,
Eqs.~(\ref{sigmaLT})--(\ref{sigmaDIP}), involve directly the
cross--sections, which are obtained from the respective amplitudes after
integrating over $b$\,: $\sigma_{\rm dipole}(r,\tau)=2\int \rmd^2b\,
T(r,b, \tau)$, etc. The unitarity constraint then becomes more subtle:
with increasing energy, a cross--section can rise indefinitely, even
after the `black disk' limit has been achieved at central impact
parameters, because of the radial expansion of the black disk. This
expansion is however quite slow, {\em logarithmic} in the energy (to
comply with Froissart bound), which is indeed much slower than the
corresponding {\em power--like} increase of the amplitude in the BFKL
regime (see below). Hence, it is possible to neglect the Froissart
expansion (or, at least, treat this in an adiabatic approximation) in a
first study of unitarization. Then, the unitarity bound reads
$\sigma_{\rm DD}(r,r',\tau)\le 2\pi r_>^2$ (with $r_> =\max(r,r')$) for
the dipole--dipole scattering, and $\sigma_{\rm dipole}(x,r)\le 2\pi
R_0^2$ for the dipole--onium one. It is in fact more suggestive to
introduce scattering amplitudes averaged over $b$, e.g., $T(r,\tau)\equiv
\sigma_{\rm dipole}(r,\tau)/(2\pi R_0^2)$, in terms of which the
unitarity bound reads simply $T\le 1$.

We can now explain our strategy for computing the saturation line:
Starting in the regime where the scattering is weak, $T\ll 1$, we compute
the (average) scattering amplitude from the solution to the BFKL equation
and then extrapolate the result towards the unitarity limit $T\sim
\order{1}$. The {\em saturation momentum} $Q_s(\tau)$ is then obtained
via the condition
  \beq\label{Tsat}
T(r,\tau)\,\simeq\,1\qquad{\rm for}\qquad
 r\,\simeq\,1/Q_s(\tau)\,.\eeq
That is,  $1/Q_s(\tau)$ is the critical dipole size for the onset of {\em
unitarity corrections} (in the form of multiple scattering) in
dipole--target scattering at rapidity $\tau$ \cite{SCALING,MT02}. This is
also the critical momentum scale for {\em parton saturation} in the
target wavefunction \cite{GLR,MQ85,MV,AM99,SAT}, in a frame where the
target has rapidity $\tau$\,: for transverse momenta $k_\perp\simle
Q_s(\tau)$, the gluon occupation numbers saturate at a value $n_g\sim
1/(\alpha_sN_c)$, which in turn implies that the (anti)quark occupation
numbers saturate at a value $n_q\sim 1$. (Recall that, at small $x$ and
relatively low $k_\perp$, the quark distribution is driven by gluons,
which dominate the small--$x$ part of the wavefunction.) This profound
link between dipole unitarization and gluon saturation in pQCD can be
understood as a consequence of the relation $T(\tau, r)\sim \abar
n_g(\tau,k_\perp\sim 1/r)$, valid in the BFKL regime, between the dipole
amplitude and the gluon occupation number.

By using Eq.~(\ref{Tsat}) together with the BFKL solutions to be shortly
presented, we shall compute the saturation lines for both dipole--dipole,
and dipole--onium, scattering. For more clarity, let us anticipate here
the main conclusions that we shall arrive at: \texttt{(i)} The value of
the saturation momentum depends upon the nature of the target, but its
evolution with increasing energy does not --- the evolution is universal.
\texttt{(ii)} The saturation momentum of the onium coincides with that of
its `softest' component --- the constituent dipole with maximal size
$r'\sim R_0$.

We now close the parenthesis and return to the calculation of the
cross--section in the BFKL approximation. When applied to the
dipole--dipole cross--section $\sigma_{\rm DD}(r,r',\tau)$, the BFKL
equation \cite{BFKL} reads  (with $\abar\equiv\alpha_sN_c/\pi$)
 \beq \frac{\del}{\del \tau}\,\sigma_{\rm DD}(r,r',\tau) &
 =&\frac{\bar{\alpha}_s}{2\pi} \int \rmd^2 {\bm z}\,
 \frac{\bm{r}^2}{\bm{z}^2 (\bm{r}-\bm{z})^2}\nn
&{}&\qquad\big\{ -\sigma_{\rm DD}(r,r',\tau) + \sigma_{\rm
DD}(z,r',\tau)+ \sigma_{\rm DD}(|\bm{r}-\bm{z}|,r',\tau)\big\}\,.
\label{BFKLhom}
 \eeq
This equation has been written here as an evolution in the projectile
size $r$ for a given target size $r'$. (Of course, the function
$\sigma_{\rm DD}(r,r',\tau)$ is symmetric in $r$ and $r'$.) It has a
simple physical interpretation: in one evolution step ($\tau \to
\tau+\rmd\tau$), the dipole $\bm{r}$ can split into two dipoles, with
sizes $\bm{z}$ and respectively $\bm{r}-\bm{z}$, which then interact with
the target. The `BFKL kernel' $\frac{\bar{\alpha}_s}{2\pi}\,
[{\bm{r}^2}/{\bm{z}^2 (\bm{r}-\bm{z})^2}]$ represents the differential
probability for such a splitting to occur per unit rapidity. To solve
Eq.~(\ref{BFKLhom}), we also need an initial condition at low energy
($\tau=0$). To the present accuracy, this is computed as the exchange of
two gluons, which yields
 \beq\label{sigmaDD0}
 \sigma_{\rm DD}(r,r',\tau=0) \,=\,2\pi\alpha_s^2 r_<^2\left(1+
 \ln\frac{r_>}{r_<}\right),\eeq
where $r_< =\min(r,r')$ and $r_> =\max(r,r')$. This cross--section shows
`color transparency' --- it vanishes like $r_<^2$ when $r_<\to 0$ ---,
meaning that a very small dipole scatters only weakly. By using
Eq.~(\ref{sigmaDD0}), one can easily check that, for $\tau=0$ and $r\ll
R_0$, the dipole--onium cross--section (\ref{sigmaDIP}) is dominated by
relatively large target dipoles, with $r \ll r'\le R_0$. (When $\Delta=
0$, the large dipoles dominates over the small ($r' < r$) ones by a large
logarithm $\ln(R_0/r)$.) As we shall later see, the feature is preserved
by the evolution with increasing $\tau$.

The solution to the BFKL equation with the initial condition
(\ref{sigmaDD0}) is most conveniently obtained in Mellin space, where it
reads
  \beq\label{TBFKL}
\sigma_{\rm DD}(r,r',\tau)\,=\, \pi\alpha_s^2 r'^2 \int\limits_{C}
\frac{\rmd\gamma}{2\pi i}\,\frac{1}{\gamma^2(1-\gamma)^2}\,
 \left(\frac{r^2}{r'^{2}}\right)^{1-\gamma}
 {\rm e}^{\tau \chi(\gamma)}
  \eeq
where $\chi(\gamma)$ is the eigenvalue of the BFKL kernel in Mellin
space, generally referred to as the {\em BFKL characteristic function}
 \beq\label{chigamma}
\chi(\gamma)= \abar\{2\psi(1)-\psi(\gamma)-\psi(1-\gamma)\},\qquad
 \psi(\gamma)\equiv \rmd \ln \Gamma(\gamma)/\rmd\gamma,\eeq
and the integration contour $C$ runs parallel to the imaginary axis with
$0< {\rm Re}(\gamma) < 1$. Note the symmetry property
$\chi(\gamma)=\chi(1-\gamma)$, which reflects the conformal invariance of
the BFKL evolution.  In view of this, one can check that
Eq.~(\ref{TBFKL}) is symmetric under $r\leftrightarrow r'$, as it should.

The inverse Mellin transform in Eq.~(\ref{TBFKL}) can be evaluated via a
saddle point approximation, in which the value of the saddle point
depends upon the balance between the kinematical variables $\abar\tau$
and $\rho-\rho'= \ln(r'^2/r^2)$. Here, we have introduced the logarithmic
variables $\rho\equiv \ln(R_0^2/r^2)$ and $\rho' \equiv \ln(R_0^2/r'^2)$
which are convenient for what follows. At this point, the reference scale
inside the logarithms is arbitrary (it anyway cancels out in the
difference $\rho-\rho'$), but for later convenience we have chosen it as
the onium size $R_0$.

Consider first the {\em formal} high--energy limit $\abar\tau\to \infty$
at fixed $\rho-\rho'$ (this is formal since the respective solution
violates the unitarity bound already for relatively low energies, as we
shall shortly see). Then the saddle point $\gamma_0$ is determined by the
minimum of the characteristic function :
 \beq
 \chi'(\gamma_0)=0\quad\Longrightarrow\quad \gamma_0=1/2\,, \label{formal} \eeq
and Eq.~(\ref{TBFKL}) can be evaluated by expanding $\chi(\gamma)$ around
$\gamma_0$, to second order (`diffusion approximation'): writing
 \beq
 \gamma=1/2-i\nu\,, \label{argu} \eeq
one obtains (notice that $\chi(1/2-i\nu)$ is an even function of $\nu$,
by conformal symmetry)
 \beq\label{diffBFKL}
 \chi(1/2-i\nu)\,=\,\omega_0 - D_0\nu^2 +
 \mathcal{O}(\nu^4)\quad\mbox{with}\quad \omega_0=4\ln 2\abar,\quad
 D_0=14\zeta(3)\abar\,,\eeq
and then the Gaussian integration over $\nu$ can be easily performed to
yield
 \beq\label{TPOMERON}
 \sigma_{\rm DD}(r,r',\tau)\,\simeq\,8\alpha_s^2
 \frac{{\rm e}^{\omega_0\tau} }{\sqrt{\pi D_0 \tau}}\,
 \left({r^2}{r'^{2}}\right)^{1/2}
  \exp\left\{-\frac{\ln^2(r^2/r'^2)}
 {4D_0\tau} \right\}.\eeq
\comment{This expression exhibits the main attributes of the `BFKL
Pomeron': the Regge--like exponential increase with $\tau$, with a
relatively large intercept $\omega_0$, the anomalous dimension 1/2, which
significantly reduces the dependence upon $r^2$ as compared to `color
transparency'\footnote{Color transparency is eventually recovered at
sufficiently large values of $\rho$ (for fixed $\tau$), i.e., for
sufficiently small dipole sizes. Indeed, when $\rho\gg\abar\tau$, the
integral in Eq.~(\ref{TBFKL}) is controlled by the `DGLAP' saddle point
at $\gamma=0$.}, and the diffusion in $\ln r^2$, which reflects the
conformal symmetry of the BFKL kernel.} This expression exhibits the
`BFKL Pomeron', i.e., an exponential increase with $\tau$ at fixed $r$
and $r'$. Clearly, when extrapolated at large $\tau$, this behaviour
violates the unitarity bound. For instance, when $r\sim r'$, this happens
at a critical rapidity $\tau_{\rm cr}\sim (1/\omega_0)\ln(1/\alpha_s^2)$,
which is parametrically not that high. (Recall that the `high--energy'
regime starts only at $\tau \sim 1/\abar$.)

\begin{figure}
\begin{center}
\includegraphics[width=12.5cm,bb=100 250 650 650]{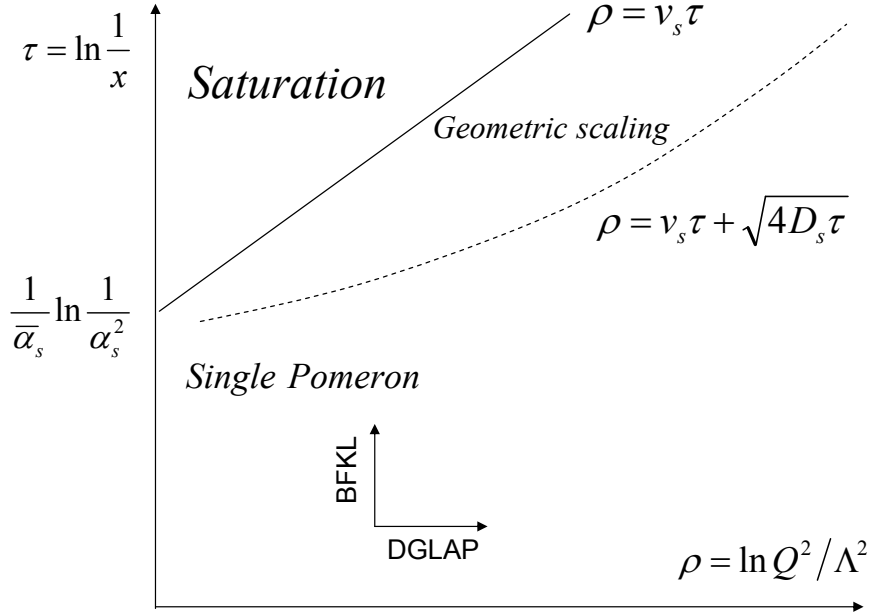}
\end{center}
\caption{\sl A `phase diagram' for the
high energy evolution in pQCD in the presence of unitarity corrections and
gluon saturation. (We recall that $\abar\equiv\alpha_sN_c/\pi$.)}
\label{FigQCD}
\end{figure}

It is nevertheless possible to explore higher rapidities $\tau>\tau_{\rm
cr}$ while staying within the limits of the BFKL approximation provided
we evolve along an {\em oblique} direction in the kinematical plane
$\rho-\tau$ (see also Fig.~\ref{FigQCD}) : when increasing $\tau$, we
should simultaneously increase $\rho$ (i.e., decrease the ratio $r/r'$),
in such a way that the scattering amplitude remains small:
$T(r,r',\tau)\ll 1$. In particular, the {\em saturation line}
$\rho=\rho_s(\tau)$ corresponds to a direction of evolution along which
the amplitude is constant and of order one, cf. Eq.~(\ref{Tsat}).

This also means that, in order to follow the saturation line, one must
abandon the symmetry between the projectile and the target dipoles
($r\leftrightarrow r'$), which has been manifest so far. If the target
dipole size is fixed to $r'$, then with increasing $\tau$ we must evolve
towards smaller and smaller values for $r$. It is then natural to write
$\sigma_{\rm DD}(r,r',\tau)=2\pi r'^2 T(r,r',\tau)$, where $r < r'$ and
the unitarity bound corresponds to $T(r,r',\tau)=1$. The BFKL
approximation holds so long as $T(r,r',\tau)\ll 1$, but it allows us to
approach the saturation line from small $r$ (large $\rho$) values, and
thus compute the saturation momentum according to Eq.~(\ref{Tsat}). To
that aim, however, one cannot rely on the Pomeron saddle point,
Eq.~(\ref{formal}), anymore, rather one must determine the saddle point
$\gamma_s$ corresponding to an evolution along the saturation line
($\rho=\rho_s(\tau)$). By combining the saddle point condition:
 \beq
  \tau \chi'(\gamma_s) = -(\rho_s-\rho')\,,
  \eeq
with the condition that the amplitude (\ref{TBFKL}) be approximately
constant along the saturation line:
 \beq
 \tau \chi(\gamma_s) - (1-\gamma_s)(\rho_s-\rho') - \ln(1/\alpha_s^2)\,=0,
 \eeq
(we have only kept the dominant parametric dependencies upon $\tau$,
$\rho_s$ and $\alpha_s$), one obtains
 \beq\label{gammasBFKL}
 \frac{\chi'(\gamma_s)}{\chi(\gamma_s)}&\,=\,&-\frac{1}{1-\gamma_s}
 \quad\Longrightarrow\quad \gamma_s\approx 0.372,\nn
 \rho_s(\tau,r')&\,=\,&\rho'+ v_s\tau -\frac{\ln(1/\alpha_s^2)}{1-\gamma_s}
  \quad\mbox{with}\quad
 v_s\,\equiv\,\frac{\chi(\gamma_s)}{1-\gamma_s}
 \,\approx 4.883\abar\,.\eeq
(We have assumed here that $\tau$, and hence $\rho_s$, are large enough
for the term $\ln(1/\alpha_s^2)$ to be treated as a small perturbation.)
Note that $\gamma_s$ is a pure number, independent of either $\tau$ or
$\abar$, corresponding to the fact that $\rho_s(\tau)$ is a {\em
straight} line, with slope $v_s\sim\order{\abar}$.

For $\rho$ larger than $\rho_s$, but not {\em much} larger, one can
estimate the amplitude by expanding Eq.~(\ref{TBFKL}) around $\gamma_s$ :
writing $\gamma=\gamma_s-i\nu$, and expanding to second order in $\nu$,
one finds
 \beq\label{expsat}
 \tau \chi(\gamma) -(1-\gamma)\rho - \ln(1/\alpha_s^2)
  \simeq - (1-\gamma_s)(\rho-\rho_s) - i\nu
 (\rho-\rho_s) - D_s\tau\nu^2,\eeq
where $D_s=\chi''(\gamma_s)/2 \approx 24.26\abar$. This limited expansion
is valid so long as $1<\rho-\rho_s\ll  D_s\tau$. In this range, the
amplitude is obtained by performing the Gaussian integration over $\nu$,
as
 \beq\label{TSAT}
 T(r,r',\tau)\,\simeq\,
 \frac{\ \  \left({r^2}{Q_s^{2}}\right)^{1-\gamma_s}
}{\sqrt{\pi D_s \tau}}\,
  \exp\left\{-\frac{\ln^2(r^2 Q_s^2)}
 {4D_s\tau} \right\},\eeq
where the {\em saturation momentum} (here, for a target dipole with size
$r'$)
   \beq\label{QSAT}
  Q_s^2(r',\tau)\,=\,(\alpha_s^2)^{1/(1-\gamma_s)}
  \,\frac{1}{r'^2}\,\rme^{v_s\tau}
  \eeq
appears as the natural reference scale for measuring the projectile
dipole size. When viewed as a function of $rQ_s$, the amplitude near
saturation, Eq.~(\ref{TSAT}), is {\em universal}, i.e., independent of
the specific properties of the target, but uniquely fixed by the
evolution. Note that, in Eq.~(\ref{TSAT}), $\gamma_s$ plays the role of
an anomalous dimension. The evolution of the saturation momentum with
increasing $\tau$, cf. Eq.~(\ref{QSAT}), is universal as well, and
controlled by the `saturation exponent' $v_s$. The above calculation
yields $v_s\sim \mathcal{O}(1)$ for $\abar= 0.2 \div 0.3$, a rather large
value that would be inconsistent with the phenomenology at HERA or RHIC.
However, it turns out that, after taking into account the
next--to--leading order BFKL corrections \cite{camici,cia}, this value is
reduced to $v_s\approx 0.3$ \cite{DT02}, which is indeed in the ballpark
of the various phenomenological analyses (see, e.g.,
\cite{GBW99,geometric,IIM03} for studies at HERA).

We now return to the problem of dipole--onium scattering, cf.
Eqs.~(\ref{sigmaDIP})--(\ref{nr}), with the purpose of computing the
corresponding saturation momentum $Q_s(\tau)$. As before, we shall
approach the saturation line from the weak scattering regime, i.e., from
relatively small dipole sizes $r \ll 1/Q_s(\tau)$. This procedure meets
with a subtle point though: although the {\em overall} scattering is, by
assumption, weak, $T(r,\tau)\ll 1$, this is not necessarily so for all
the individual dipoles which compose the onium and which contribute to
the convolution in Eq.~(\ref{sigmaDIP}). The problem comes from the
relatively small dipoles which, according to Eq.~(\ref{QSAT}), develop
large saturation momenta (meaning that they evolve into `hot spots' with
high gluon density), off which the projectile dipole can strongly
scatter. Hence, when evaluating the contribution of such small dipoles to
the total cross--section, one cannot rely on the BFKL approximation
anymore. Still, as we shall shortly argue, this brings no serious
complication, since, after being properly unitarized, the small dipoles
give negligible contributions to the overall cross--section in the
vicinity of the saturation line.


To see this, one can use the following, piecewise, approximation for the
dipole--dipole cross--section at high energy,
 \beq\label{sigmaDDtau}
    \sigma_{\rm DD}(r,r',\tau)\,
     \approx\,2\pi r_>^2
    \begin{cases}
        \displaystyle{T\big(r_<\,Q_s(r_>,\tau),\,\tau\big)}
         &
        \text{ for\,  $r_<\,\ll\, 1/Q_s(r_>,\tau)$}
        \\*[0.2cm]
        \displaystyle{1} &
        \text{ for\,  $r_<\,\simge\, 1/Q_s(r_>,\tau)$}
        \,.
    \end{cases}
    \eeq
In the first line, $T(r_<\,Q_s(r_>,\tau),\tau)$ is the amplitude in the
weak scattering regime, as given by Eqs.~(\ref{TSAT})--(\ref{QSAT}) after
replacing $r \to r_<$ and $r' \to r_>$. The second line represents the
black disk regime. The precise transition between these two regimes is
not under control, but this is irrelevant for the present purposes. By
inserting Eq.~(\ref{sigmaDDtau}) into Eq.~(\ref{sigmaDIP}) and performing
the integration over $r'$, one can estimate the dipole--onium scattering
amplitude $T(r,\tau) \equiv {\sigma_{\rm dipole}(r,\tau)}/{2\pi R_0^2}$.
The very small target dipoles with $r' < r$ are easily seen to be
irrelevant, since their contribution to $T(r,\tau)$ is of order
$r^2/R_0^2 \ll 1$. (Recall that we are only interested in contributions
which can approach the unitarity limit $T=1$ when $r\to 1/Q_s(\tau)$.) To
evaluate the contribution of the larger dipoles with $r < r' \le R_0$, it
is again convenient to use logarithmic variables ($\rho\equiv
\ln(R_0^2/r^2)$, etc.), and notice that Eq.~(\ref{gammasBFKL}) can be
rewritten as
 \beq\label{rhos0}
 \rho_{s}(\tau,r')\,=\,\rho'+ \rho_{s0}(\tau),\qquad
 \rho_{s0}(\tau)\,\equiv\, v_s\tau
 -\frac{\ln(1/\alpha_s^2)}{1-\gamma_s}\,,
 \eeq
where  $\rho_{s0}(\tau)$ represents the saturation momentum for a dipole
with the maximal size $r'=R_0$. We anticipate here that in the regime of
interest we have $\rho > \rho_{s0}(\tau)$. Then the contribution of the
target dipoles with $\rho' < \rho$ (i.e., $r' > r$) can be evaluated as
  \beq\label{sigmaDO}
  T(\rho,\tau)
  \,\simeq\,\int\limits_0^{\rho-\rho_{s0}}\rmd\rho'
  \,{\rm e}^{-(1-\gamma_s)(\rho-\rho'-\rho_{s0})}\,
 {\rm e}^{-\frac{(\rho-\rho'-\rho_{s0})^2}
 {4D_s \tau} } \,{\rm e}^{-(1+\Delta)\rho'}
 \,+\, \int\limits_{\rho-\rho_{s0}}^\rho
 \rmd\rho'  \,{\rm e}^{-(1+\Delta)\rho'}\,,
 \eeq
where the first (second) term in the r.h.s. corresponds to large dipoles
which scatters only weakly (respectively, to relatively small dipoles for
which the unitarity limit has been reached). The first integral in the
r.h.s. involves the overall exponential factor
 \beq \label{over}
 {\rm e}^{(1-\gamma_s)\rho'}\,{\rm e}^{-(1+\Delta)\rho'}\,=\,
 \,{\rm e}^{-(\Delta+\gamma_s)\rho'}\,\eeq
where the exponent $\Delta+\gamma_s$ is strictly positive in pQCD (recall
that $\Delta\ge 0$ and $\gamma_s\approx 0.37$). It is then easy to check
that, so long as $\rho-\rho_{s0} < 2(\Delta+\gamma_s) D_s\tau$, this
integral is dominated by its lower limit $\rho'=0$, i.e., by target
dipoles of the largest possible size $r'\sim R_0$. As for the second
integral in Eq.~(\ref{sigmaDO}), this is dominated by its own lower
limit, at $\rho'= \rho-\rho_{s0}$. One thus finds
  \beq\label{sigmaDO1}
  T(\rho,\tau)
  \,\sim\,\frac{1}{\Delta+\gamma_s}\,
  {\rm e}^{-(1-\gamma_s)(\rho-\rho_{s0})}\,
 {\rm e}^{-\frac{(\rho-\rho_{s0})^2}
 {4D_s \tau} }
 \,+\, \frac{1}{\Delta+1}  \,{\rm e}^{-(1+\Delta)(\rho-\rho_{s0})}\,.
 \eeq
As just mentioned, $\Delta+\gamma_s>0$, hence the second term in the
r.h.s. is exponentially suppressed w.r.t. the first one. Thus, finally
(up to a slowly varying prefactor),
  \beq\label{sigmaDOFIN}
  T(\rho,\tau)
  \,\sim\,
  {\rm e}^{-(1-\gamma_s)(\rho-\rho_{s0})}\,
 {\rm e}^{-\frac{(\rho-\rho_{s0})^2}
 {4D_s \tau} }\qquad\mbox{for}\qquad
 \rho-\rho_{s0} < 2(\Delta+\gamma_s) D_s\tau\,,\eeq
which is essentially the same result as for a target made with a single
dipole of size $R_0$ (compare to Eq.~(\ref{TSAT})). In particular, the
saturation momentum of the onium coincides with that of its largest
dipole component (cf. Eq.~(\ref{QSAT})) : $\rho_s(\tau)=\rho_{s0}(\tau)$,
or
 \beq\label{QSATO}
  Q_s^2(\tau)\,=\,(\alpha_s^2)^{1/(1-\gamma_s)}
  \,\frac{1}{R_0^2}\,\rme^{v_s\tau}\,.
  \eeq
This is quite intuitive: the onium starts to look `black' as a whole only
when a large dipole, which covers all (or most) of the hadron disk, has
evolved into a system with high gluon density on the resolution scale of
the projectile. Smaller dipoles reach the unitarity limit much faster (on
that particular resolution scale), but their contributions to the total
cross--section are suppressed by their small area\footnote{Of course, the
physical picture would be different if $\Delta$ was negative and large
enough, in such a way to bias the dipole distribution (\ref{nr}) towards
small sizes. Then, the onium could become black as a whole (on a given
resolution scale) by getting covered with many small dipoles which are
individually black.}.

The high--energy dipole amplitude in Eqs.~(\ref{TSAT}) or
(\ref{sigmaDOFIN}) exhibits an interesting structure, which can be better
appreciated by comparison with the low--energy amplitude
(\ref{sigmaDD0}), or the (purely) BFKL prediction at high energy,
Eq.~(\ref{TPOMERON}): Besides the power--law $r^{2(1-\gamma_s)}$ with
anomalous dimension $\gamma_s\approx 0.37$, there is also a Gaussian
factor describing {\em diffusion} in the logarithmic variable $\rho\sim
\ln(1/r^2)$. This shows that, when increasing $\tau$, the amplitude gets
built via a diffusive process which gets most of its support from the
region $0 <\rho-\rho_s <\sqrt {4 D_s\tau}$ above the saturation line. For
$\tau$ large enough, such that $Q_s^2(\tau)\gg\Lam^2$, this region lies
fully in the perturbative domain, thus confirming the internal
consistency of our present approach. The physics of
saturation/unitarization eliminates the symmetric diffusion
characteristic of the (pure) BFKL evolution, cf. Eq.~(\ref{TPOMERON}),
which would invalidate the use of perturbation theory.

In particular, within the more restricted window $\rho-\rho_s\ll \sqrt{4
D_s\tau}$ (close to the saturation line), the diffusion term in
Eq.~(\ref{sigmaDOFIN}) can be ignored, and then the amplitude shows {\em
geometric scaling}, i.e., it depends upon $r$ and $\tau$ only via the
single variable $rQ_s(\tau)$ :
 \beq\label{TGEOM}
 T(r,\tau)\,\sim\,
 {\rm e}^{-(1-\gamma_s) (\rho-\rho_s)}\, =\,
 \left({r^2}{Q_s^{2}(\tau)}\right)^{1-\gamma_s}
 \qquad\mbox{for}\qquad
 \rho-\rho_{s}\ll \sqrt{4 D_s\tau}
 \,.\eeq
This scaling property is an important consequence of the non--linear
dynamics responsible for unitarization on the shape of the amplitude in
the weak scattering regime `above saturation' ($\rho > \rho_s$). Via the
dipole factorization (\ref{sigmaLT}), this property gets transmitted to
the DIS cross--section in the high $Q^2$ regime at $Q^2 > Q_s^{2}(\tau)$
:
 \beq \frac{F_2(Q^2,\tau)}{Q^2}\,\sim\, \sigma_{\gamma^*p}(Q^2,\tau)\sim
 N_c R_0^2\left(\frac{Q_s^2(\tau)}{Q^2}\right)^{1-\gamma_s}
 \quad\mbox{for}\quad Q_s^{2}(\tau)\simle Q^2\ll
 Q_g^{2}(\tau), \label{f2}
 \eeq
where $Q_g^{2}(\tau)\equiv Q_s^{2}(\tau)\rme^{\sqrt{4 D_s\tau}}$ is the
upper bound of the scaling window, cf. Eq.~(\ref{TGEOM}).

It is furthermore interesting to note the behaviour of $F_2$ for very
small values of $Q^2$, deeply at saturation ($Q^2\ll Q_s^{2}(\tau)$).
Then, the virtual photon cross--section is dominated by large dipoles,
with $1/Q_s^{2}(\tau) < r^2 < 1/Q^2$, for which the black disk limit is
saturated: $\sigma_{\rm dipole}(r,\tau)\approx {2\pi R_0^2}$. One then
finds (the logarithm is generated by the integral over $r^2$)
 \beq \frac{F_2(Q^2,\tau)}{Q^2}\,\sim\,
 N_c R_0^2\ln\frac{Q_s^2(\tau)}{Q^2}\qquad\mbox{for}\qquad Q^2\ll
 Q_s^{2}(\tau), \label{f2sat}
 \eeq
which exhibits geometric scaling too. Remarkably, it turns out that a
similar scaling has been observed indeed \cite{geometric} in the HERA
data for DIS at small $x\le 0.01$ and up to relatively large values of
$Q^2\sim 100$ GeV$^2$ (well above the saturation momentum at HERA,
estimated in the ballpark of 1 GeV). The `phase--diagram' for high energy
evolution in pQCD which emerges from the previous considerations is
graphically summarized in Fig.~\ref{FigQCD}.

\section{DIS at strong coupling}

\subsection{The setup}

\setcounter{equation}{0}

With this section we begin the study of the most interesting problem for
us here, namely DIS at strong 't Hooft coupling and high energy.
Specifically, our physical gauge theory will be a deformation of the
${\mathcal N}=4$ SYM theory in the regime where the gauge coupling is
weak, $g\ll 1$, but the number of colors is large, $N_c\gg 1$, in such a
way that the `t Hooft coupling $\lambda\equiv g^2N_c$ (the relevant
coupling for perturbation theory at large $N_c$) is large : $\lambda \gg
1$. The `deformation' refers to the introduction of an effective infrared
cutoff which breaks down conformal symmetry and mimics confinement.

According to the AdS/CFT correspondence \cite{malda}, the ${\mathcal
N}=4$ SYM theory is admitted to have a dual description in terms of the
Type IIB string theory living in the 10--dimensional $AdS_5 \times S^5$
space--time, with the following metric:
 \beq \label{met}
 \rmd s^2=\frac{u^2}{R^2}(-\rmd x_0^2+\rmd x_1^2+\rmd x_2^2+\rmd x_3^2)+
 \frac{R^2}{u^2}\rmd u^2 +R^2\rmd \Omega^2_5\,. \eeq
In this equation, $x_0$ and $\bm{x}=(x_1,x_2,x_3)$ are the time and,
respectively, three spatial dimensions of `our' physical Minkowski world,
$u$ is the $AdS_5$ radial coordinate ($0\le u <\infty$), to be below
referred as `the 5th dimension', and $\rmd\Omega^2_5$ is the angular
measure on $S^5$. Furthermore, $R$ is the common radius of $AdS_5$ and
$S^5$, and it is related to the string length $l_s$ via the `t Hooft
coupling of the dual gauge field theory: $R=l_s\lambda^{\frac{1}{4}}=
\alpha'^{\frac{1}{2}}\lambda^{\frac{1}{4}}$, with $\alpha'\equiv l_s^2$
the Regge slope. The physical, 4--dimensional, space is identified with
the boundary of $AdS_5$ at $u\to\infty$, where the $AdS_5$--metric
becomes conformal to the standard Minkowski metric $\eta^{\mu\nu}$. The
string coupling constant $g_s$ and the gauge coupling $g$ are related to
each other as $g_s=g^2/4\pi =\lambda/4\pi N_c$, which shows that in the
`strong coupling' regime of interest here ($\lambda\gg 1$ but $g\ll 1$),
the string theory is weakly coupled and hence it can be treated in a
semi--classical approximation.

We have previously mentioned the `deformation' of the ${\mathcal N}=4$
SYM theory which is necessary to mimic confinement. In the dual string
theory, this translates into a modification  of the metric at small
$u\sim u_0\equiv \Lambda R^2$, where $\Lambda$ is a parameter which sets
the scale of light hadrons (`glueballs'). The precise way of modifying
the geometry in the infrared is an issue of debate which we do not enter.
(As we shall see, the dominant behaviour at high energy is insensitive to
such details.) In what follows, we shall simply restrict the 5th
coordinate to values $u\ge u_0$ (in practice, we shall be mostly
interested in $u\gg u_0$), and assume that the spectrum of the
supergravity theory includes states which are localized near $u=u_0$,
corresponding to massive `glueballs' on the gauge theory side. Note that,
because of the logarithmic nature of the metric in $u$ ($\rmd s^2_u
\propto \rmd u^2/u^2$), massive `supergravity' states exhibit power--law
decays at large $u$, to be better characterized later on.

To closely mimic the setup of DIS, we follow \cite{dis} and introduce a
`photon' as the gauge boson associated with a gauged $U(1)$ subgroup of
the $SU(4)$ ${\mathcal R}$-symmetry. As for the target, we employ a
`supergravity' glueball of mass $\sim \Lambda$ which originates from the
Kaluza--Klein decomposition of the 10 dimensional massless dilaton after
compactification on $S^5$. Still as in Ref. \cite{dis}, we shall use
indices $M,N,...$ to denote all ten space--time dimensions, separating
into $\mu,\nu,...$ on the Minkowski boundary at $u\to \infty$ (our
physical space), $m,n,...$ on $AdS_5$, and $a,b,...$ on $S^5$. Indices
are raised or lowered in the standard way, using the metric $G_{MN}$
which can be read off Eq.~(\ref{met}) : $\rmd s^2=G_{MN} \rmd x^M \rmd
x^N$. It is understood that for quantities living in the
four--dimensional gauge theory (and hence also on the Minkowski boundary
of $AdS_5$), the tensor operations are performed with the Minkowski
metric $\eta^{\mu\nu}=(-1,1,1,1)$.

As standard in DIS, the structure functions are obtained from the
imaginary part of the forward Compton scattering amplitude
 \beq
 i\int \rmd^4x\, \rme^{iq\cdot x}\langle
 P|T\{J^\mu(x)J^\nu(0)\}|P\rangle\,,
  \eeq
where $J^\mu(x)$ is the density of the ${\mathcal R}$-current and
$|P\rangle$ denotes a (normalizable) glueball state with 4--momentum
$P^\mu$. The hadronic matrix element under the integral will be computed
from semi--classical string theory, via the AdS/CFT correspondence. The
recipe is as follows :

The ${\mathcal R}$-current at the Minkowski boundary ($u\to\infty$)
excites vector--like metric fluctuations in the `bulk' (i.e., at finite
$u\ge u_0$) :
  \beq \delta G_{ma}(x_\mu,u,\Omega)\,=\,A_m(x_\mu,u) v_a(\Omega),\eeq
where $\Omega$ denotes the angular coordinates on $S^5$, $A_m(x_\mu,u)$
is a gauge boson field, and $v_a$ is the Killing vector on $S^5$
corresponding to the gauged $U(1)$ subgroup. At $u\to\infty$, the field
$A_\mu(x_\mu,\infty)$ can be identified as the physical photon which
couples to the ${\mathcal R}$-current. This will be taken in the form
$A_\mu(x_\mu,\infty)=n_\mu \rme^{iq\cdot x}$, with space--like
four--vector $q^\mu$ : $Q^2\equiv  q^\mu q_\mu > 0$. But inside the bulk,
$A_m(x_\mu,u)$ should be better viewed as a gravitational wave, as it
describes a fluctuation in the metric tensor. For definiteness, we shall
refer to this field as the `${\mathcal R}$-boson'. This ${\mathcal
R}$-boson propagates into the bulk until it scatters off the dilaton
target --- a supergravity state with wavefunction
  \beq \Psi(x_\mu,u,\Omega)\,=\, \rme^{iP\cdot x}\,\Phi(u,\Omega).\eeq
Note that this is a plane wave with four--momentum $P^\mu$ in the four
physical dimensions. The Bjorken--$x$ variable for DIS is defined in the
usual way: $x\equiv Q^2/(2P\cdot q)$. Still as usual, the high--energy
limit ($P\to\infty$ at fixed $Q^2$) corresponds to small values for $x$
($x\ll 1$).

As explained in Refs. \cite{dis}, the Regge behavior starts to be seen
when $\,1/x \gg \sqrt{\lambda}\,$ where the energy is high enough to
create excited string intermediate states. To leading order at large
$N_c$, the string scattering can be computed at tree level,  as the
imaginary part of the superstring Virasoro--Shapiro amplitude for
graviton--dilaton scattering \cite{Polchinski} (originally calculated in
flat space, and heuristically extended to curved space in \cite{dis})
convoluted with the supergravity wavefunctions for the ${\mathcal
R}$-boson ($A_m$) and the dilaton ($\Phi$). The result of \cite{dis} can
be compactly written as
 \beq \label{st} F_2(x,Q^2)= c \alpha'\,\frac{
(QR)^6}{\Lambda^2} \int \frac{\rmd
 u}{u}\frac{1}{u^4}\left(K_0^2(QR^2/u)+K_1^2(QR^2/u)\right)
 \left(\frac{1}{x}\right)^{1+\frac{\alpha' \Delta_2}{2}} \Phi^\dagger
 \Phi(u)\,, \nonumber \\ \eeq
where we have absorbed in $c\,$ a constant coming from the integration
over $S^5$, as well as powers of $\lambda$. (For simplicity, we do not
keep track of powers of $\lambda$ in the prefactor. This is acceptable
since we shall deal with an exponentially large factor
$\,\rme^{\sqrt{\lambda}}$. See below for details.)

Let us discuss the various factors under the integral in Eq.~(\ref{st})
one after the other:

\texttt{(i)} The modified Bessel functions $K_{0,1}$ are a part of the
`${\mathcal R}$-boson wavefunction'; that is, they arise via the solution
$A_m(x_\mu,u)$ to the supergravity equation of motion for the metric
perturbation induced by the ${\mathcal R}$-current. This equation is
Maxwell equation on $AdS_5$: $D^mF_{mn}=0$, with $F_{mn}=\partial_m
A_n-\partial_n A_m$ and $D^m$ the covariant derivative in the $AdS_5$
background. ($A_m(x_\mu,u)$ is the non--normalizable solution to this
equation corresponding to the boundary condition
$A_\mu(x_\mu,u\to\infty)=n_\mu \rme^{iq\cdot x}$.) For $u$ smaller than
 \beq u_c \equiv QR^2,
\label{uc}
 \eeq
the Bessel functions decays exponentially, meaning that a `photon' with
large $Q^2$ cannot penetrate deeply inside the bulk: the larger $Q^2$ is,
the closer the photon remains near the boundary. Since, on the other
hand, the inverse power of $u$ manifest in Eq.~(\ref{st}) favors small
values of $u$, it is clear that the integral there is controlled by
$u\sim u_c$.

Remarkably, the ${\mathcal R}$-boson wavefunction involves the same
Bessel functions as the wavefunction describing the $q\bar q$ excitation
of a virtual photon in pQCD (cf. Eq.~(\ref{PsiTL})). This suggests a kind
of `duality' between the dipole size $r$ in pQCD and the 5th dimension
$u$ in $AdS_5$, via the correspondence $r\leftrightarrow R^2/u$. We shall
identify further aspects of this correspondence later on.

\texttt{(ii)} The dilaton wavefunction $\Phi(u)$ is peaked around $u\sim
u_0$ and decays at large $u$ like $\Phi\sim u^{-\Delta}$, where $\Delta$
is the conformal dimension of the operator which creates this state. For
glueballs corresponding to supergravity modes, $\Delta$ is a number of
${\mathcal O}(1)$ which is strictly larger than 2. In the formal analogy
to pQCD, the probability density $|\Phi(u)|^2\sim u^{-2\Delta}$
corresponds to the density $n(r)$ of dipoles inside the onium, cf.
Eq.~(\ref{nr}). Note that the overlap between the ${\mathcal R}$-boson
and the dilaton wavefunction scales like $u_c^{-2\Delta}\sim
(1/Q^{2})^\Delta$, and hence it rapidly dies away when increasing $Q^2$.


\texttt{(iii)} The last comment shows that DIS would be strongly
suppressed at large $Q^2$ if there was not for the non--locality hidden
in the middle factor, $(1/x)^{1+(\alpha'/2) \Delta_2}$, which describes
the imaginary part associated with one string exchange in the curved,
$AdS_5$, space. In less formal notations, the action of this operator can
be written as (with $\tau\equiv\ln(1/x)$)
 \beq\label{DeltaAction}
 \left(\frac{1}{x}\right)^{1+\frac{\alpha' \Delta_2}{2}} \Phi^\dagger
 \Phi(u)\,\equiv\,\frac{1}{x}\,\int\rmd u'\, \langle u|
 \rme^{\frac{\alpha' \tau\Delta_2}{2}}|u'\rangle\,
 |\Phi(u')|^2\,.\eeq
Notice the dominant energy behavior $F_2(x) \sim 1/x$ at small $x$: this
is the hallmark of the graviton exchange in the $t$--channel. (At high
energy, the exchanged string is dominated by its graviton component; see
Sect. 4.2.) Furthermore, the operator $\Delta_2$ in the exponent is the
Laplacian describing diffusion in the 5th dimension,  for the spin--2
exchanged graviton. This operator appears because of the finite momentum
transfer in the $u$--direction, although the momentum transfer in the
physical four dimensions is strictly zero \cite{dis,pomeron}. Explicitly,
this operator reads
 \beq  \label{delta2} \Delta_2\equiv
 \left(\frac{u}{R}\right)^2
  \nabla_0^2 \left(\frac{u}{R}\right)^{-2} = \frac{4}{R^2}(\partial^2_\rho-1)\eeq
where $\nabla_0^2$ is the scalar Laplacian on $AdS_5$ and we have defined
$\rho \equiv \ln (u^2/u_0^2)$. By also using the integral representation
of the `heat kernel' $\exp\{\partial^2_\rho\}$, one arrives at the
following representation for the expression in Eq.~(\ref{DeltaAction}) :
 \beq \label{last} \rme^{\tau+2\alpha'\tau
(\partial^2_\rho-1)}\, \Phi^\dagger \Phi(\rho)&=&\int \rmd \rho' \int
\frac{\rmd \nu}{2\pi}\ \rme^{i\nu(\rho-\rho')}\
\rme^{\tau-\frac{2\tau}{\sqrt{\lambda}}(\nu^2+1)}\ |\Phi(\rho')|^2
\nonumber \\ \nonumber\\
  &{=}&\ \frac{\rme^{\omega_0 \tau}}{2\sqrt{\pi D\tau}}
  \int_0^\infty \rmd \rho'\
 \rme^{\frac{-(\rho-\rho')^2}{4D\tau}}\ \rme^{-\Delta \rho'},
 \eeq
where we have used $R^2/\alpha' = \sqrt{\lambda}$ and introduced
 \beq
 \omega_0 \equiv 1-\frac{2}{\sqrt{\lambda}}\,,\qquad\mbox{and}\qquad
 D\equiv \frac{2}{\sqrt{\lambda}}\,.\eeq
Eq.~(\ref{last}) result shows a Pomeron--like increase with $\tau$, with
an intercept $1+\omega_0=2-2/\sqrt{\lambda}$ which is slightly shifted
from 2 due to the curvature of $AdS_5$ \cite{hard,dis,pomeron}.

One should notice the formal similarity between the stringy `Pomeron'
amplitude in Eq.~(\ref{last}) and the cross--section (\ref{TPOMERON})
which describes the BFKL Pomeron in pQCD. In this analogy, the fifth
coordinate $u$ plays, once again, the role of the inverse dipole size
$1/r$. Like for the BFKL Pomeron, the non--locality associated with the
single Pomeron exchange in $AdS_5$ is diffusive in $\rho\sim \ln u^2$,
with the rapidity $\tau$ playing the role of the evolution `time';
accordingly, its effect becomes important only for very high energies,
{\em exponentially} large in the relevant coupling. Namely, the diffusive
radius $D\tau$ in Eq.~(\ref{last}) becomes of $\order{1}$ when $\tau \sim
\sqrt{\lambda}$, or $1/x \sim \rme^{\sqrt{\lambda}}\,$. For comparison,
the corresponding estimate in pQCD reads $\tau \sim 1/\abar$, or $1/x
\sim \rme^{1/\abar}$, which at a first sight may look as an
astronomically high energy, but in fact is is not:  $\abar$ is not that
small in real world ($\abar=0.2\div 0.4$ for the high--energy
experiments), and rapidities of order $1/\abar$ or even larger are within
the reach of the present--day accelerators.

Up to an overall factor of $\order{1/N_c^2}$ that we shall shortly
comment on, Eq.~(\ref{last}) can be viewed as the string theory analog of
the dipole--onium scattering amplitude in pQCD, in the BFKL
approximation. The dipole (a $q\bar q$ excitation of the virtual photon
with a given transverse size $r$) is now replaced by the ${\mathcal
R}$-boson $A_m(u)$ (the metric fluctuation induced by the ${\mathcal
R}$-current at a given position $u$ along the 5th dimension), and the
dilaton plays the same role as the onium --- the corresponding
wavefunctions are localized near the respective `infrared' cutoff ($u\sim
u_0$ for the dilaton and, respectively, $r\sim R_0$ for the onium) and
they exhibit power--law tails with exponent $\Delta$ in the `ultraviolet'
(large $u$ and, respectively, small $r$). At this stage, it would be
straightforward to repeat the analysis in Sect. 2 ({\em mutatis
mutandis}) and thus determine the saturation momentum for the dilaton.
However, before doing that, it is necessary to better understand the
validity limits of the approximation in Eq.~(\ref{last}). We shall do
that in Sect. 3.2, where we shall discover that, in the string theory
context, Eq.~(\ref{last}) is less general than its analog in pQCD --- it
holds only within a limited range of values for $\tau$ and $\rho$. These
limitations have not been properly recognized in the previous literature.

We conclude this subsection with a discussion of the interplay between
the large--$N_c$ limit and the high--energy limit, which turns out to be
quite subtle, for both weak and strong coupling. The tree--level string
amplitude (corresponding to single Pomeron exchange) is of order
$g_s^2\propto 1/N_c^2$. In Eq.~(\ref{st}) this factor $1/N_c^2$ has been
canceled by a factor $N_c^2$ implicit in the ${\mathcal R}$-current. (The
latter receives contributions from all the fields in the ${\mathcal N}=4$
SYM theory which are charged under the ${\mathcal R}$-symmetry; these
fields are scalars and Weyl fermions in the adjoint representation of the
colour group, hence they have $N_c^2-1\approx N_c^2$ degrees of freedom.)

This is similar to what happens in QCD at weak coupling: the BFKL result
(\ref{TBFKL}) involves an overall factor $\alpha_s^2\sim
\bar{\alpha}_s^2/N_c^2$ which is partially canceled in $F_2$ by a factor
$N_c$ coming from the photon wavefunction, Eq.~(\ref{PsiTL}). Yet, the
unitarity corrections apply directly to the dipole--dipole scattering
(and not only to $F_2$), hence they become important when the small
factor $1/N_c^2$ manifest in Eqs.~(\ref{TBFKL}) or (\ref{TPOMERON}) gets
compensated by the energy evolution. As discussed below
Eq.~(\ref{TPOMERON}), for the BFKL Pomeron this happens at a critical
rapidity $\tau_{\rm cr} \sim (1/\bar{\alpha}_s) \ln (N_c^2/\abar^2)$. For
$\tau \simge \tau_{\rm cr}$, multiple Pomeron exchanges become important
and ensure unitarization.

Returning to strong coupling and the dual string theory, there the
multiple Pomeron exchanges correspond to higher genus string amplitudes
which are suppressed by powers of $1/N_c^2$. Hence, unitarity corrections
become important at $\tau_{\rm cr} \sim \ln N_c^2$. We thus see that, in
the strict large--$N_c$ limit ($N_c\to \infty$ at fixed energy), there is
no issue of unitarity: higher loop amplitudes, being proportional to
inverse powers of $1/N_c^2$, can be made arbitrarily small even at very
high energy. On the other hand, when $N_c$ is large but {\em finite}, the
single Pomeron exchange has only a limited applicability, and beyond that
one has to sum over an infinite number of multi--loop diagrams
corresponding to arbitrarily many Pomerons exchanged in the $t$--channel.
What is however different from the situation at weak coupling is that,
for generic values of $\tau$ and $\rho\sim \ln Q^2$, the amplitude in the
vicinity of the unitarity line is not dominated by the single Pomeron
exchange anymore. This is why, in general, the saturation line cannot be
inferred from the previous formul\ae{} in this section, but requires some
further analysis.

It will be our goal in what follows to determine the onset of the various
types of unitarity corrections in the parameter space of  $\tau$ and
$\rho$. To have access to the full structure of the high--energy `phase
diagram', we shall assume that $N_c^2 > \rme^{\sqrt{\lambda}}$ --- so
that diffusion effects (which, we recall, require a rapidity evolution
$\tau \simge \sqrt{\lambda}$) become important before unitarization. But
the opposite case $ \rme^{\sqrt{\lambda}}\ll N_c^2$ will be included too
in our results, as a special limit.

\subsection{Limit of the single Pomeron approximation :
A lesson from a sum rule}

The fundamental difference between the physics of unitarization at weak
and, respectively, strong coupling comes from the fact that, in the
strongly--coupled gauge theories with a gravity dual, the high--energy
scattering amplitude is predominantly real (it starts as one graviton
exchange), while at weak coupling it is predominantly imaginary (in both
QCD and ${\mathcal N}=4$ SYM theory). Of course, the large real part at
strong coupling is irrelevant for the total cross section so long as the
latter is computed at {\em tree--level}. But at {\em higher loop} level,
say, starting with two--graviton exchange, this gives rise to an
imaginary part via {\em diffractive} processes --- i.e., processes where
the final states (string excitations or higher Kaluza--Klein modes) are
separated by a rapidity gap. On the other hand, the imaginary part of the
tree--level amplitude (\ref{st}) is associated with the {\em inelastic}
production of excited string states, with no rapidity
gap\footnote{Strictly speaking, the DIS process is always inelastic,
since the virtual photon cannot appear in the final state. Here, we use
the word `inelastic' to suggestively distinguish the final states without
rapidity gap from the diffractive ones, which exhibit such a gap.}. It
turns out that, when $\rho\sim \ln Q^2$ is larger than a certain critical
value, the diffractive processes dominate over the inelastic ones in the
approach towards unitarity. We shall discuss this issue in detail in the
next section, but for the time being let us proceed naively and study the
onset of unitarity corrections on the basis of the tree--level results in
the previous subsection, which include the inelastic processes alone.
This will drive us into a paradox which will point out towards the
necessity to include the additional, diffractive, contributions.

After also including the essential factor $1/N_c^2$, the forward
scattering amplitude (\ref{last}) for the ${\mathcal R}$-boson--dilaton
can be more suggestively rewritten as (compare to Eq.~(\ref{sigmaDIP}))
 \beq\label{GDT}
 T(\rho,\tau)\,=\,\int_0^\infty \rmd \rho'\
  T_0(\rho,\rho',\tau)\ |\Phi(\rho')|^2\,
 \eeq
where $|\Phi(\rho')|^2=\rme^{-\Delta \rho'}$ and
 \beq\label{GGT}
 T_0(\rho,\rho',\tau)\,\equiv\,\frac{1}{N_c^2}\,
 \frac{\rme^{\omega_0 \tau}}{2\sqrt{\pi D\tau}}\
 \rme^{-\frac{(\rho-\rho')^2}{4D\tau}} \,
 \eeq
represents the elementary scattering amplitude between states localized
around $\rho$ and $\rho'$ in the single-Pomeron-exchange approximation
(the analog of the dipole--dipole amplitude in Sect. 2). The unitarity
bound on this elementary amplitude determines the saturation line for a
dilaton state localized at $\rho'$; namely, the condition
$T_0(\rho,\rho',\tau)\sim 1$ for $\rho=\rho_s(\tau,\rho')$ implies
 \beq\label{rhosGD0}
 \rho_{s}(\tau,\rho')\,=\,\rho'+ \rho_{s0}(\tau),\qquad
 \rho_{s0}(\tau)\,\equiv\, \sqrt{4D\tau(\omega_0 \tau-\ln
N_c^2)}\,,
 \eeq
where  $\rho_{s0}(\tau)$ refers to a dilaton state localized at $u\sim
u_0$. (As in Sect. 2, we neglect the effects of the slowly varying
prefactor $1/(2\sqrt{\pi D\tau})$ in Eq.~(\ref{GGT}).)

Eq.~(\ref{rhosGD0}) applies for $\tau > \tau_{\rm cr}= (1/\omega_0)\ln
N_c^2$ and should be compared to the corresponding result,
Eq.~(\ref{rhos0}), in pQCD: the similarity between these two results
would become even closer if it was possible to extrapolate
Eq.~(\ref{rhosGD0}) to large rapidities $\tau \gg \tau_{\rm cr}$, where
$\ln N_c^2$ could be expanded out from the square root. However, as we
shall later discover, such an extrapolation would be incorrect: in
reality, Eq.~(\ref{rhosGD0}) applies only within a limited range in
$\tau$ above $\tau_{\rm cr}$, namely, within the window $\tau_{\rm
cr}<\tau <\tau_c$, with $\tau_c -\tau_{\rm cr}\simeq {2\ln
N_c^2/\sqrt{\lambda}}$ (see Eq.~(\ref{inter}) below).

This limitation is in turn related to a failure of our formula
(\ref{GGT}) for the elementary amplitude, which becomes incomplete at
high energies and for large values of $\rho$. To illustrate this failure,
we shall show that Eq.~(\ref{GGT}) leads to an unphysical result when
used to evaluate the following moment of the structure function
 \beq\label{tauint}
 M(Q^2)\equiv \int_0^1 \rmd x \, F_2(x,Q^2)=\int_0^\infty \rmd \tau\,
 \rme^{-\tau}F_2(\tau,Q^2)\,.
 \eeq
By energy--momentum conservation, this moment should be independent of
$Q^2$; however, as we shall demonstrate in what follows, the (properly
unitarized) amplitude for single Pomeron exchange, Eq.~(\ref{GGT}), fails
to saturate the sum--rule at large values of $Q^2$.

Within the present approximations, the structure function $F_2$ of the
dilaton is given by Eq.~(\ref{st}). To avoid unnecessary complications
with the ${\mathcal R}$-boson wavefunction, we shall here restrict
ourselves to the structure function that would be measured by a single
${\mathcal R}$-boson state localized at $u$ (the analog of a single
dipole projectile with fixed size $r$ in pQCD). The coordinate $u$ fixes
the resolution scale, $u^2\propto Q^2$ (cf. Eq.~(\ref{uc})), hence
$F_2(\tau,Q^2)$ can be evaluated as
 \beq\label{F2u}
 F_2(\tau,Q^2)\ \sim \ N_c^2\,\rme^{\rho}\,T(\rho,\tau)\,
 \eeq
(up to an irrelevant prefactor), with $\rho \equiv \ln (u^2/u_0^2)$. The
factor of $N_c^2$ accounts for the color degrees of freedom of the fields
contributing to the ${\mathcal R}$-current and $\rme^{\rho}$ accounts for
the factor $Q^2$ in the relation (\ref{fff2}) between $F_2$ and the
`dipole' (here, ${\mathcal R}$-boson) cross--section.

We start by evaluating $T(\rho,\tau)$ in the single-Pomeron
approximation. This is given by Eq.~(\ref{GDT}) with the elementary
amplitude from Eq.~(\ref{GGT}) {\em corrected for unitarity violations}.
That is, for $\tau > \tau_{\rm cr}$, we shall write (compare to
Eq.~(\ref{sigmaDO}) in pQCD)
 \beq\label{TGD0}
  T(\rho,\tau)
  \,\simeq\,\int\limits_0^{\rho-\rho_{s0}}\rmd\rho'
  \,\frac{1}{N_c^2}\
 {\rme^{\omega_0 \tau}}\,
 \rme^{-\frac{(\rho-\rho')^2}{4D\tau}} \,{\rm e}^{-\Delta\rho'}
 \,+\, \int\limits_{\rho-\rho_{s0}}^\infty
 \rmd\rho'  \,{\rm e}^{-\Delta\rho'}\,.
 \eeq
As in Sect. 2, this representation is valid only in the regime where the
amplitude is weak, $T(\rho,\tau)\ll 1$, but it can be used to approach
the unitarity limit $T\sim 1$, and thus compute the dilaton saturation
momentum. The first integral in the r.h.s. of Eq.~(\ref{TGD0}) involves
the dilaton components living at relatively small values of $u$, which
scatter only weakly off the incoming ${\mathcal R}$-boson. The second
integral refers to the components at larger values of $u$, which undergo
strong scattering, but are exponentially suppressed by the dilaton
wavefunction. (Of course, when $\tau \le \tau_{\rm cr}$ only the first
integral would be present.) Accordingly, the overall amplitude is
dominated by the components living near the infrared cutoff, so like in
pQCD.

Indeed, the integrand in the first integral is peaked at $\rho'=\rho-
2\Delta D\tau$; hence, for $\rho < 2\Delta D \tau$, that integral is
dominated by its lower limit $\rho'=0$, which yields
\beq\label{TGD1}
  T(\rho,\tau)
  \,\sim\,
  \frac{1}{N_c^2}\
 {\rme^{\omega_0 \tau}}\,
 \rme^{-\frac{\rho^2}{4D\tau}}
 \,+\, \frac{1}{\Delta}  \,{\rm e}^{-\Delta(\rho-\rho_{s0})}\,.
 \eeq
This amplitude becomes of $\order{1}$ when $\rho\simeq\rho_{s0}(\tau)$,
meaning that the saturation momentum of the dilaton coincides with that
of its component located at $u\simeq u_0$ :
$\rho_{s}(\tau)=\rho_{s0}(\tau)$. However, the above calculation of
$\rho_s$ is consistent only so long as $\rho_s < 2\Delta D \tau$, meaning
$\rho_{s0}< 2\Delta D \tau$, which in turn requires $\tau<\tau_d$, with
 \beq\label{taud}  \tau_d\,\equiv\,\frac{\ln
 N_c^2}{\omega_0-D\Delta^2}\,=\,\frac{\ln
 N_c^2}{1-(2/\sqrt{\lambda})(\Delta^2+1)}\,. \eeq
When $\tau<\tau_d$, one can check that the second term in
Eq.~(\ref{TGD1}) is exponentially suppressed next to the first one, and
therefore
 \beq\label{TGDfin}
  T(\rho,\tau)
  \,\simeq\,
  \frac{1}{N_c^2}\
 {\rme^{\omega_0 \tau}}\,
 \rme^{-\frac{\rho^2}{4D\tau}}
 \qquad\mbox{when} \quad
 \tau<\tau_d\quad\mbox{and}\quad
 \rho < 2\Delta D \tau\,.
 \eeq
Thus, in this regime, the scattering is controlled by the component of
the dilaton living near the IR cutoff $u_0=\Lambda R^2$, in full analogy
to what happens in pQCD (recall the discussion after
Eq.~(\ref{sigmaDO1})). In particular, within the restricted interval
$\tau_{\rm cr} < \tau<\tau_d$, Eq.~(\ref{TGDfin}) can be rewritten in a
form which resembles the expansion of the QCD amplitude near the
saturation line (compare to Eq.~(\ref{sigmaDOFIN}))
 \beq\label{TGDscaling}
  T(\rho,\tau)
  \,\simeq\,
  {\rm e}^{-(1-\gamma)(\rho-\rho_{s})}\,
 {\rm e}^{-\frac{(\rho-\rho_{s})^2}
 {4D \tau} }\qquad\mbox{when} \quad \tau_{\rm cr} <
 \tau<\tau_d\quad\mbox{and}\quad
 \rho_{s}< \rho < 2\Delta D \tau\,,\nn
 \eeq
with the `anomalous dimension' $\gamma(\tau)$ given by
 \beq\label{gamma0}
 1-\gamma(\tau)\,=\,\frac{\rho_{s0}(\tau)}{2D\tau}\,=\,
 \sqrt{\frac{\omega_0 \tau-\ln
 N_c^2}{D\tau}}\,.\eeq
Note however that, even for $\rho$ close to $\rho_{s}$, where the
diffusion term can be ignored in Eq.~(\ref{TGDscaling}), this amplitude
does {\em not} show geometric scaling (unlike in pQCD), because the
$\tau$--dependence of the `anomalous dimension' (\ref{gamma0}) cannot be
neglected within the corresponding validity range. This difference can be
understood as follows: in pQCD, expressions like Eq.~(\ref{sigmaDOFIN})
are valid up to arbitrarily large values of $\tau$, much larger than the
critical value $\tau_{\rm cr}\sim (1/\abar)\ln(1/\alpha_s^2)$ for the
onset of saturation; then properties like geometric scaling emerge via
asymptotic expansions valid at $\tau\gg \tau_{\rm cr}$. By contrast, in
the present context, Eq.~(\ref{TGDscaling}) is valid only within a
limited range of values for $\tau$ above $\tau_{\rm cr}$, where such
asymptotic expansions are not possible anymore.


The case where $\rho > 2\Delta D \tau$ can be similarly treated. For
$\tau<\tau_d$, the saddle point $\rho'=\rho- 2\Delta D\tau$ lies within
the integration range and gives the dominant contribution to the overall
amplitude:
 \beq\label{TGD2}
  T(\rho,\tau)
  \,\simeq\,
  \frac{1}{N_c^2}\
 {\rme^{(\omega_0 +D\Delta^2)\tau - \Delta\rho}}\,
 \qquad\mbox{when} \quad
 \tau<\tau_d\quad\mbox{and}\quad
 \rho > 2\Delta D \tau\,,
 \eeq
in agreement with a corresponding result in Ref. \cite{dis}. For
$\tau>\tau_d$, Eq.~(\ref{TGD0}) would predict $T(\rho,\tau)\sim {\rm
e}^{-\Delta(\rho-\rho_{s0})}$, but this result cannot be trusted anymore
since, as we shall shortly argue, Eq.~(\ref{TGD0}) is not a reliable
approximation for such large values of $\rho$ and $\tau$.

To see that, consider the prediction of Eq.~(\ref{TGD0}) for the
sum--rule (\ref{tauint}). When computing the integral there, the
amplitude (\ref{TGD0}) can be replaced by its simpler expression in
Eq.~(\ref{TGDfin}) (this will be checked later). To also account for the
unitarity bound, we shall write
 \beq\label{TSP}
  T(\rho,\tau)
  \,\simeq\,
      \begin{cases}
        \displaystyle{\frac{1}{N_c^2}\
  {\rme^{\omega_0 \tau}}\,
  \rme^{-\frac{\rho^2}{4D\tau}}
   }      &
        \text{ for\,  $\tau \,<\,\tau_s(\rho)$}
        \\*[0.2cm]
        \displaystyle{1} &
        \text{ for\,  $\tau\,\simge\, \tau_s(\rho)$}
        \,.
    \end{cases}
    \eeq
Here, $\tau_s(\rho)$ is the dilaton saturation line expressed as a
function of $\rho$ (the `inverse' of the function $\rho_{s}(\tau)$); that
is, $\tau_s(\rho)$ is the rapidity at which the unitarity corrections
become important on the resolution scale fixed by $\rho$. Using
Eq.~(\ref{rhosGD0}), one immediately finds
 \beq \tau_s(\rho)\,=\,\frac{\ln N_c^2}{2\omega_0}\,+\,
 \sqrt{\left(\frac{\ln N_c^2}{2\omega_0}\right)^2\,+\,
 \frac{\rho^2}{4D\omega_0}}.
   \label{taus} \eeq
After using (\ref{TSP}) together with the estimate (\ref{F2u}) for $F_2$,
the integral in Eq.~(\ref{tauint}) becomes
\beq\label{Mint}
 M(\rho)\ \sim \ {\rme^{\rho}}\,
 \int_0^{\tau_s} \rmd \tau\,
 \rme^{-(1-\omega_0)\tau}\, \rme^{-\frac{\rho^2}{4D\tau}}\ +
 \ N_c^2\,{\rme^{\rho}}\,
 \int_{\tau_s}^\infty \rmd \tau \,\rme^{-\tau} \,\equiv\,M_1+M_2\,.
 \eeq
The second integral is trivially performed, and its result is found to
rapidly vanish at large $\rho$:
  \beq\label{Isat}
 M_2(\rho)\,\sim\,N_c^2\,{\rme^{\rho}}\,\rme^{-\tau_s(\rho)}
 \, \longrightarrow \, \exp\left\{-\frac{\lambda^{1/4}}{\sqrt{8}}
 \,\rho\right\}
 \quad\mbox{for}\quad \rho\to\infty.
 \eeq
Hence, the only way to satisfy the sum--rule would be via the first
integral, $M_1$. The corresponding integrand is strongly peaked near the
saddle point at
  \beq \tau^*(\rho)\,=\,\frac{\sqrt{\lambda}}{4}\,\rho\,.
   \label{gra} \eeq
Note that, for $\tau\sim\tau^*$, we have $2\Delta D \tau\sim \Delta \rho
> \rho$, which justifies our use of the simplified expression
(\ref{TGDfin}) for the single--Pomeron amplitude in Eq.~(\ref{TGD0}).

So long as $\tau^*(\rho) < \tau_s(\rho)$, this saddle point lies inside
the integration domain and can be used to estimate the integral. The
ensuing result for $M_1(\rho)$ is independent of $\rho$, showing that the
sum--rule is saturated by the single--Pomeron exchange, as it should for
this to be a consistent approximation. This is the situation considered
in Ref. \cite{dis}, within the framework of the strict large--$N_c$
limit. Note, however, that in that limit, the saturation rapidity
$\tau_s$ was effectively pushed up to infinity and unitarity has never
been an issue\footnote{This is also the situation in pQCD, where the
sum--rule (\ref{tauint}) is dominated by
relatively large values of $x\sim\order{1}$, 
hence its calculation is insensitive to unitarity corrections, or even to
the BFKL evolution.}.

However, when $N_c$ is kept finite (although large), unitarity {\em is}
an issue, which modifies, in particular, the previous calculation of
$M_1(\rho)$ at large values of $\rho$. Namely, the function
$\tau^*(\rho)$ grows faster than $\tau_s(\rho)$, so the two curves cross
each other at a point (see also Fig.~\ref{phase2})
 \beq (\tau_c,
 \rho_c)=\left(\, \frac{\ln N_c^2}{1-4/\sqrt{\lambda}}\,, \frac{4\ln
 N_c^2}{\sqrt{\lambda}-4}\, \right) \,. \label{inter} \eeq
Hence, when $\rho> \rho_c$, the saddle point (\ref{gra}) lies outside the
integration range. Then, the integral giving $M_1$ is dominated by its
upper cutoff at $\tau=\tau_s(\rho)$, which however yields only a tiny
result, of the same order as $M_2$ (since for $\tau\sim\tau_s(\rho)$, the
amplitude (\ref{TSP}) is $\order{1}$).


Thus, for $\rho> \rho_c$ the sum--rule is not saturated by the single
Pomeron exchange anymore, meaning that some new physical mechanism, which
was missed by this approximation, should become important in this region.
The above calculation gives us some indications in that respect: We have
already noticed that, when increasing $\rho$ towards $\rho_c$, the saddle
point (\ref{gra}) approaches the saturation line (\ref{taus}), that it
crosses exactly at $\rho= \rho_c$. It is then reasonable to assume that,
for $\rho> \rho_c$, the saddle point should remain along the saturation
line: $\tau^*(\rho) =\tau_s(\rho)$. (One cannot have $\tau^*(\rho) >
\tau_s(\rho)$, since the integral over $\tau$ in the saturation region is
always dominated by its lower limit $\tau_s(\rho)$, as manifest on
Eq.~(\ref{Mint}).) If so, then the sum--rule at large $\rho$ can be
estimated as in Eq.~(\ref{Isat}): $M(\rho)\sim
N_c^2\,\exp\{\rho-\tau_s(\rho)\}$. This estimate, together with the
condition $M(\rho)=\order{1}$, allows us to predict the form of the
saturation line at large $\rho> \rho_c$\,:
 \beq\label{tausat}
 \tau_s(\rho) \,\simeq\,\rho+\ln N_c^2\,,\qquad\mbox{or}\qquad
 \rho_s(\tau)\,\simeq\,\tau - \ln N_c^2\,.\eeq
This discussion suggests that the complete saturation line should
actually involve two branches, given by Eq.~(\ref{taus}) for $\rho<
\rho_c$ and, respectively, Eq.~(\ref{tausat}) for $\rho> \rho_c$.
Remarkably, these two branches appear to continuously match with each
other at the transition point (\ref{inter}) : both their values and their
first derivatives coincide at this point (see Fig.~\ref{phase2}).

What should be the unitarization mechanism responsible for the saturation
line in Eq.~(\ref{tausat})~? Once again, the previous sum--rule
calculation gives us an indication in that sense: Along the line which
controls the sum--rule, the energy dependence of the ${\mathcal
R}$-boson--dilaton amplitude is precisely that of the massless graviton
(\emph{not} Pomeron) exchange:
\beq\label{massless} {\rme^{\omega_0 \tau}}\
 \rme^{\frac{-\rho^2}{4D\tau}} \,=\,
 \rme^{(1-4/\sqrt{\lambda})\tau} \,=\,
 {\rme^{\tau}}\  {\rme^{-\rho}}\,.
 \eeq
(We have repeatedly used $\tau=({\sqrt{\lambda}}/{4})\rho$, cf.
Eq.~(\ref{gra}).) Eq.~(\ref{massless}) is indeed the same as the
propagator of a massless graviton. This result is quite suggestive : The
shift of the Pomeron intercept $1\to \omega_0$ in Eq.~(\ref{last}) can be
understood as a result of gravitons acquiring a mass due to the curvature
of $AdS_5$ (see Sect. 4.2 below for details). However, along the line
Eq.~(\ref{gra}), and in fact in the whole region on the right of this
line as we shall see, the presence of massless gravitons is inevitable.
The graviton exchange is dominantly real and much more non--local than
the Pomeron exchange, which is non--local only due to diffusion. However,
double, or multiple, graviton exchanges can generate imaginary parts, via
the diffractive final states alluded to at the beginning of this
subsection. As we shall explicitly see later on, such multiple exchanges
are indeed responsible for unitarization at $\rho> \rho_c$ and they
produce the saturation line in Eq.~(\ref{tausat}).

\section{Mapping the high--energy `phase diagram'}

Motivated by the above considerations, in this section we shall study the
Pomeron propagator in the $AdS_5$ geometry, with due attention to both
its real and its imaginary part. This will allow us to follow the
transition between a genuine Pomeron behaviour at relatively low values
of $\rho$, where the curvature effects are important and give rise to an
imaginary part and to diffusion, and a massless graviton propagator at
large $\rho$, which is long--ranged and predominantly real. Then we shall
explain how multiple graviton exchanges generate an imaginary part via
diffractive processes, and compute the corresponding saturation line. As
a preliminary, we shall make some comments on the anomalous dimension of
twist--two operators at strong coupling, which justify the use of the
`diffusion approximation' in the approach towards the saturation line.

\subsection{Anomalous dimension of twist--two operators in ${\mathcal N=4}$ SYM}
\setcounter{equation}{0}

The variable $\gamma$ introduced in Sect.~2 has the interpretation as the
anomalous dimension of the twist--two, spin--$j$ gluonic operator
\beq
 \Tr[F_\mu^+(D^+)^{j-2}F^{+\mu}], \label{two}
 \eeq
with $j$ analytically continued to non--integer values. (In the  BFKL
context one has $j=1+\chi(\gamma)$; see \cite{lev} for discussions on the
anomalous dimension and the operator product expansion in that
framework.) Recall that the total dimension $\Delta(j)$ of the operator
Eq.~(\ref{two}) is given by\footnote{This is the common definition of
$\gamma$ in the small--$x$ literature, which differs by a factor $-2$
from the one which is most often used in the literature. A yet different
convention $\gamma \to 1-\gamma$ is sometimes used.}
  \beq \Delta(j)\,=\,j+2-2\gamma(j)\,. \label{dim} \eeq
In the weak coupling  expansion, $\gamma \sim {\mathcal O}(\alpha_s)$,
while the BFKL resummation gives a number of ${\mathcal O}(1)$ (and
comprised between 0 and 1/2; cf. Sect. 2). On the other hand, for
${\mathcal N}=4$ SYM at strong `t Hooft coupling $\lambda=g^2N_c \gg 1$,
the AdS/CFT correspondence predicts that the twist--two operators with
$j\neq 2$ receive a large, negative (in our conventions), anomalous
dimension. Specifically, Ref.~\cite{gubser} has found
\beq \gamma(j)
\approx -\frac{\sqrt{\lambda}}{2\pi}\ln \frac{j}{\sqrt{\lambda}}\, ,
\label{cusp}
 \eeq
for $j\gg \sqrt{\lambda}$ and, respectively,
\beq \gamma(j) \approx
-\sqrt{\frac{j}{2}}\,\lambda^{\frac{1}{4}}\, , \label{jj} \eeq for $1 \ll
j \ll \sqrt{\lambda}$. These results were obtained by computing the
energy and the angular momentum of a `folded' closed string state
rotating in the $AdS_5$ space. When $j\ll \sqrt{\lambda}$, the string is
spinning in a small region compared with the radius $R$, while the case
$j \gg \sqrt{\lambda}$ corresponds to a macroscopic string stretched over
a distance larger than $R$.

Because of the large anomalous dimension for twist--two operators
(excepting the energy momentum tensor which has zero anomalous dimension
$\gamma(2)=0$), it has been argued in \cite{dis} that the OPE in DIS at
strong coupling must be considerably reorganized. The dominant
contribution comes from the `protected' (i.e., non--renormalized),
double--trace, operators which create and annihilate the entire hadron of
interest. These operators are nominally higher--twist, but since their
dimension is of order unity, their contribution is less suppressed
compared to the twist--two operators having $|\gamma(j)| \gg 1$ for
generic $j\neq 2$. Hence DIS at moderate values of $x$ can be pictured as
a process where a virtual photon knocks off an entire charged hadron -- a
drastically different picture from pQCD where a photon knocks off partons
inside the hadron.

While plausible in the Bjorken limit ($Q^2\to \infty$ at fixed $\tau$),
this is no longer the case in the Regge limit ($\tau\to \infty$ at fixed
$Q^2$). The important regime at high energy is $j\sim 2$. Since
$\gamma(2)=0$ exactly, $\gamma(j)$ remains small  in the vicinity of
$j=2$, and the twist--two  operators (analytically continued to complex
values of $j$ near $j=2$) should still give the leading contribution.
Notably, Refs.~\cite{kotikov} and \cite{pomeron} arrived at the same
expression for  $\gamma(j)$ in the `diffusion approximation' from very
different arguments. It is obtained as the solution to
\beq \label{chara}
  j-j_0 =\frac{1}{2\sqrt{\lambda}}(\Delta-2)^2=
  \frac{1}{2\sqrt{\lambda}}(j-2\gamma(j))^2\,,
   \eeq
   where
   \beq
 j_0 = 2-\frac{2}{\sqrt{\lambda}}=1+\omega_0\,, \label{j0} \eeq
is the Pomeron intercept. Notice that the property $\gamma(2)=0$ is
indeed verified by the solution to Eq.~(\ref{chara}). Some light on the
origin of Eq.~(\ref{chara}) will be shed by the manipulations in the next
subsection. The inverse function $j(\gamma)$ is approximately quadratic
in $\gamma$ and has a minimum at
  \beq \gamma_0=1-\frac{1}{\sqrt{\lambda}}\,, \eeq
with the minimum value $j(\gamma_0)=j_0$.

In general, the diffusion approximation is expected to be valid only in
the vicinity of this minimum, i.e., when $\gamma\sim \gamma_0$, $j\sim
j_0$. Away from $j=2$, one could {\em a priori} expect significant
nonlinear corrections to the right hand side of Eq.~(\ref{chara}). This
is precisely what happens at weak coupling, in the context of the BFKL
equation, where the diffusion approximation corresponding to
$\gamma_0=1/2$, cf. Eq.~(\ref{diffBFKL}), significantly deviates from the
exact BFKL characteristic function already when $\gamma=\gamma_s\simeq
0.37$. (This is why, in that context, it would be incorrect to compute
the saturation line by `unitarizing' the amplitude (\ref{TPOMERON})
obtained via the diffusion approximation).

However, it turns out that at strong coupling the diffusion approximation
is much more robust and has a much wider validity range than at weak
coupling. For instance, one can see that Eq.~(\ref{chara}) correctly
reproduces the leading behavior, Eq.~(\ref{jj}), for large (but not
asymptotically large) $j$. (As a matter of fact, Ref.~\cite{kotikov} has
determined the proportionality constant in Eq.~(\ref{chara}) in such a
way to match Eq.~(\ref{jj}). But the derivation of Ref.~\cite{pomeron}
does not rely on this matching.) Moreover, Eq.~(\ref{chara}) coincides
with the well--known anomalous dimension formula of a scalar field in
$AdS_5$
   \beq\label{Delta}
   \Delta=2+\sqrt{4+m^2R^2}\,, \eeq
where $m$ is the mass of the state lying on the leading Regge trajectory
(i.e., the highest spin state in a given excitation level
$n=0,1,2,\dots$) :
   \beq \label{leading}
 m^2=\frac{4n}{\alpha'}=\frac{2(j-2)}{\alpha'}\,. \eeq
This shows that, in this strong--coupling context, the quadratic relation
$j\sim \gamma^2$ (`diffusion approximation') has its roots in the Regge
behavior of string excited states $\Delta^2 \sim m^2 \sim j$, which is
expected to hold in $AdS_5$ for generic values of $j$ up to $j\sim
\sqrt{\lambda}$ (see Eq.~(\ref{jj})). This range is large enough to cover
all values of $j$ and $\gamma$ of interest at high energy, as we shall
later verify.

\subsection{The Pomeron propagator}

In Eq.~(\ref{st}), the Virasoro--Shapiro formula in flat space was used
rather heuristically \cite{dis}. For a more coherent discussion on both
the real and the imaginary parts of the amplitude, it is important to
carefully derive this amplitude from the fundamental $S$--matrix in
string theory, starting with the operator representation of the latter in
curved spacetime. Up to a factor $g_s^2$ to be introduced in the final
result, this representation reads \cite{Polchinski,pomeron}
 \beq 
 \int \rmd^2z \,\langle {\mathcal V}(\infty){\mathcal
 V}(1){\mathcal W}(z){\mathcal W}(0)\rangle \approx \int_{|z|< 1} \rmd^2z
 \,\left\langle {\mathcal V}(\infty){\mathcal
 V}(1)z^{L_0-2}\bar{z}^{\tilde{L}_0-2}{\mathcal W}(1){\mathcal
 W}(0)\right\rangle \nonumber \\ =\left\langle {\mathcal V}(\infty){\mathcal
 V}(1)\frac{\delta_{L_0,\tilde{L_0}}}{L_0+\tilde{L}_0-2}{\mathcal
 W}(1){\mathcal W}(0)\right\rangle. \label{38} \eeq
Here, ${\mathcal V}\,$'s and ${\mathcal W}\,$'s are the vertex operators
corresponding to dilatons and, respectively, gauge bosons in the $(0,0)$
picture, and the integration is over the string world--sheet with
spherical topology. The average is understood in the sense of the string
path integral, which involves an integral over the zero modes and a path
integral over the nonzero ones. Furthermore, $L_0$ and $\tilde{L}_0$ are
the right-- and left--moving Virasoro operators which generate the scale
transformations on the world--sheet. The kernel
${\delta_{L_0,\tilde{L_0}}}/({L_0+\tilde{L}_0-2})$ in the last expression
can be recognized as the propagator of the closed string exchanged in the
$t$--channel.

One can diagonalize the operators $L_0$ and $\tilde{L}_0$ by inserting a
complete set of string states --- the `Pomeron states' ---, which are the
states exchanged in the $t$--channel at high--energy  \cite{pomeron}. To
that aim, one inserts the identity as a sum over the respective
projection operators
 \beq\label{proj} 1=\sum_j \int \frac{ \rmd
 \nu}{2\pi} |\,{\mathcal V}^-_{j,-\nu}\rangle \langle{\mathcal V}^+_{j,\nu}|
 \longrightarrow \int_C \frac{\rmd j}{2 i} \frac{1+\rme^{-i\pi j}}
 {\sin \pi j}\int \frac{\rmd \nu }{2\pi}\
 |\,{\mathcal V}^-_{j,-\nu}\rangle \langle{\mathcal V}^+_{j,\nu}|\,, \eeq
where the sum runs over all the positive, even, values of the spin :
$j=2,4,...$, corresponding to the string states on the leading Regge
trajectory in Eq.~(\ref{leading}). Alternatively, the sum over $j$ can be
realized as an integration in the complex $j$--plane, as indicated in the
r.h.s.; the contour $C$ encircles the positive part of the real axis and
is conveniently deformed into a contour $L$ which runs parallel to the
imaginary axis, on the left of all the poles at $j>0$ (see Fig.
\ref{contour} below). (More precisely, this contour deformation within
the Sommerfeld--Watson representation is truly an analytic continuation
from the $t$--channel amplitude related to the physical, $s$--channel,
amplitude via crossing symmetry ($t\leftrightarrow s$); see, e.g.,
\cite{gri}.)

In the equation above, ${\mathcal V}^\pm$ is the `Pomeron vertex
operator' \cite{pomeron} which is dual to the twist--two operator,
Eq.~(\ref{two}), of the gauge theory (see also \cite{tseytlin}). For
large $u$, it takes the form
 \beq
  {\mathcal V}^\pm_{j,\nu} \sim U^{j-2-2i\nu} (\partial_z
 X^\pm\partial_{\bar{z}}X^\pm)^{\frac{j}{2}} \,. \label{ver} \eeq
Capital letters denote world--sheet fields, while lower case letters
refer to their zero modes.  Also, light--coordinates are defined in the
standard way: $X^\pm=(X^0\pm X^3)/\sqrt{2}$, with the collision axis
taken along $X^3$. For what follows it is useful to notice that the
vertex operators associated with the dilaton (${\mathcal V}$) and the
${\mathcal R}$-boson (${\mathcal W}$) admit representations similar to
that in Eq.~(\ref{ver}), but with the $U$-factor replaced by the
corresponding wavefunction --- $\Psi(U)$ in the case of the dilaton and,
respectively, $A_m(U)$ for the ${\mathcal R}$-boson.

The operator (\ref{ver}) has total dimension $\Delta=2+2i\nu$; together
with Eq.~(\ref{dim}), this implies the following anomalous dimension :
\beq\label{gammaP}
 \gamma=\frac{j}{2}-i\nu\,, \eeq
which appears to be different from the one usually associated with a
leading--twist operator in the diffusion approximation, cf.
Eq.~(\ref{argu}). Let us open here a parenthesis to explain this
discrepancy: in the leading logarithmic approximation used throughout
Sect. 2, the reference scale of the logarithm $\tau=\ln s$ is ambiguous
and usually taken to be a product of the two external scales. This
ambiguity is resolved in the next--to--leading log approximation (NLLA)
where $\gamma$ is shifted in the following way \cite{camici}
 \beq
 \left(\frac{s}{Q\Lambda}\right)^{j-1}\left(\frac{Q^2}{\Lambda^2}
  \right)^{\frac{1}{2}-i\nu}
  =\left(\frac{s}{Q^2}\right)^{j-1}
  \left(\frac{Q^2}{\Lambda^2}\right)^{\frac{j}{2}-i\nu}\,,
  \eeq
to ensure that the Pomeron behaviour $({s}/{Q^2})^{j-1}$ matches with the
definition of the Bjorken--$x$ variable in DIS (at high energy).

The action of $L_0$ on the Pomeron vertex operator has been evaluated for
$j\approx 2$ \cite{pomeron} as
  \beq\label{LP} L_0{\mathcal
  V}^\pm_{j,\nu}= \tilde{L}_0{\mathcal
  V}^\pm_{j,\nu}=\left(\frac{j}{2}-\frac{\alpha}{4}\Delta_j^2\right){\mathcal
  V}^\pm_{j,\nu}= \left(\frac{j}{2}+\frac{\nu^2+1}{\sqrt{\lambda}}\right){\mathcal
  V}^\pm_{j,\nu}\,, \eeq where (c.f. Eq.~(\ref{delta2}))
  \beq \Delta_j\equiv
  \left(\frac{U}{R}\right)^j
  \nabla_0^2 \left(\frac{U}{R}\right)^{-j}\,\eeq
The $\nu$--integral (or the $j$--integral) sets $L_0=\tilde{L_0}=1$,
which after also using Eqs.~(\ref{gammaP})--(\ref{LP}) is recognized as
the previous condition (\ref{chara}). So, Eq.~(\ref{chara}) simply states
that ${\mathcal V}^{\pm}$ is a (1,1) tensor on the world--sheet
in the curved background of $AdS_5$ \cite{callan}.

We are now prepared to evaluate the string amplitude in Eq.~(\ref{38}) in
the semi--classical approximation. After using Eqs.~(\ref{proj}) and
(\ref{LP}), one finds
 \beq
&{}& \left\langle {\mathcal V}(\infty){\mathcal
 V}(1)\frac{\delta_{L_0,\tilde{L_0}}}{L_0+\tilde{L}_0-2}{\mathcal
 W}(1){\mathcal W}(0)\right\rangle = \nn\nn
 &{}&\qquad\quad =\int
 \frac{\rmd j \,\rmd \nu}{4\pi i }\,\frac{1+\rme^{-i\pi j}}{\sin \pi j}
 \, \frac{1}{j-j_0+\frac{2\nu^2}{\sqrt{\lambda}}}
 \,\langle {\mathcal V}(\infty){\mathcal
 V}(1){\mathcal V}^-_{j,-\nu}\rangle \langle {\mathcal
 V}^+_{j,\nu}{\mathcal W}(1){\mathcal W}(0)\rangle\,
 . \label{38bis} \eeq
When evaluating expectation values in the semi--classical limit, one can
replace an operator $F(U)$ on the world--sheet with its value $F(u)$ at
position $u$ (the zero mode $U$). Besides, the path integral over $U$
reduces to the ordinary integral over $u$, with the appropriate,
invariant, measure. Since the string propagator in Eq.~(\ref{38bis}) is
non--local in $u$, some care is needed when constructing the measure.
Namely, in the decomposition of unity, Eq.~(\ref{proj}), we need to
insert the vertex operators at different positions $u$ and $u'$. The
appropriate decomposition reads then
 \beq\label{proju}
 1\,=\,
 \int_L \frac{\rmd j}{2 i} \frac{1+\rme^{-i\pi j}}
 {\sin \pi j}\int \frac{\rmd \nu }{2\pi}
 \int \frac{\rmd u'}{R}
 \,\sqrt{-G'}\,(G'^{+-})^{j}
 \ |\,{\mathcal V}^-_{j,-\nu}(u')\rangle \langle{\mathcal V}^+_{j,\nu}(u)|
 \,, \eeq
where $G$ is the determinant of the $AdS_5$ metric $G_{mn}$, and $G^{+-}$
is the respective metric component;  Eq.~(\ref{met}) implies $G^{+-}=
R^2/u^2$ and $\sqrt{-G} = (u/R)^3$. Eq.~(\ref{proju}) follows from the
following decomposition of the $\delta$--function, which can be easily
checked:
 \beq 1=\int \rmd
 u'\, \delta(u-u')= \int \frac{\rmd u'}{R}
 \sqrt{-G'}(G'^{+-})^{j}\int \frac{\rmd \nu}
 {2\pi}\left(\frac{u'}{R}\right)^{j-2+2i\nu}
 \left(\frac{u}{R}\right)^{j-2-2i\nu}\,. \eeq
One can now insert the explicit expressions for the vertex operators in
the $u$--representation and then perform the integral over $\nu$ by
deforming the contour in the complex plane. Using the fact that $u > u'$
in DIS at large $Q^2$ (recall that $u\sim Q$, whereas $u'\sim u_0$), we
shall close the contour in the lower half plane, and thus pick up the
pole at
 \beq\label{nuj}
 \nu(j) \,=\,-i\sqrt{\frac{\sqrt{\lambda}}{2}\,(j-j_0)}.
 \eeq
This gives
 \begin{align} \label{las} & \int \rmd u \sqrt{-G}\int \frac{\rmd j}{2 i}
\frac{1+\rme^{-i\pi j}}{\sin \pi j}
\frac{(\alpha'_{\rm eff}s)^j}{ \sqrt{j-j_0}}
 \nonumber \\ & \qquad \times \int \rmd u'
\sqrt{-G'}(G'^{+-})^{j} \Phi^\dagger\Phi(u')
u'^{j-2+2i\nu(j)}u^{j-2-2i\nu(j)}
  A_m(u,q)A^m(u,-q)+\cdots\,,
 \nonumber \\*[0.2cm]
 & =  \int \frac{\rmd j}{2 i }\frac{1+\rme^{-i\pi j}}{\sin \pi j}
 \frac{\rme^{\tau j}}
 {\sqrt{j-j_0}}
 \int \frac{\rmd u}{u} u^4
 \int \frac{\rmd u'}{u'}  \Phi^\dagger\Phi(u')
\left(\frac{u'^2}{u^2}\right)^{1-\gamma(j)} A_m(u,q)A^m(u,-q)+\cdots\,,
\end{align}
where the dots stay for terms which are subleading at high energy. The
factor (cf. Eq.~(\ref{uc}))
   \beq
 (\alpha'_{\rm eff}s)^j= \left(\frac{\alpha'R^2s}{u^2}\right)^{j}\sim
  \left(\frac{\alpha'R^2s}{u^2_c}\right)^{j} =
  \left(\frac{1}{\sqrt{\lambda}x}\right)^j
  \sim \rme^{\tau j}\,, \eeq
arises after contraction with vertex operators which carry large momenta
\cite{pomeron}. Apart from the external wavefunctions, Eq.~(\ref{las}) is
symmetric under the exchange of $u$ and $u'$ as it should.

After including the factor $g_s^2 \sim 1/N_c^2$ and substituting the
solution of the Maxwell equation for $A_m(u)$ \cite{dis}, one can
identify the forward scattering amplitude for the current--dilaton
scattering, whose imaginary part yields the $F_2$ structure function:
$F_2 \propto  N_c^2 Q^2\,{\rm Im}\,\tilde T$. [We use the notation
$\tilde T$ for the complete scattering amplitude, including both the real
and the imaginary part; hence, $T={\rm Im}\,\tilde T$.] One finds
 \beq\label{factori} \tilde T(x,Q^2)
 &\,\simeq\,& \frac{(QR)^6}{N_c^2}
 \int \frac{\rmd u}{u}
 \frac{1}{u^4}\left(K_0^2(QR^2/u)+K_1^2(QR^2/u)\right)
 \nn &{}&\quad\times\int \frac{\rmd u'}{u'}\,  \Phi^\dagger\Phi(u')
\int\frac{\rmd j}{2 i }\,\frac{1+\rme^{-i\pi j}}{\sin \pi j}
 \,\frac{\rme^{\tau (j-1)}}
 {\sqrt{j-j_0}}\left(\frac{u'^2}{u^2}\right)^{1-\gamma(j)}.
 \eeq
Eq.~(\ref{factori}) exhibits `radial factorization': it is expressed as a
convolution in $u$ of pieces which can be recognized as wavefunctions for
the ${\mathcal R}$-boson and, respectively, the dilaton, and a kernel
which is by construction the Pomeron propagator in string theory --- a
sum over all string states exchanged in the $t$--channel at high energy
---, and which after multiplication by $g_s^2$ plays the role of a
scattering amplitude between states localized at $u$ and $u'$. Clearly,
this $u$--factorization is the string counterpart of the
`$k_T$--factorization' valid in the high--energy regime at weak coupling
(as exhibited, e.g., in Eqs.~(\ref{sigmaLT}) and (\ref{sigmaDIP})).

Let us concentrate on the elementary Pomeron amplitude, that we would
like to compare to its previous approximation in Eq.~(\ref{GGT}). With
our usual notation $\rho \equiv \ln (u^2/u_0^2)$, this can be rewritten
as
 \beq\label{SST}
 T_\mathbb{P}(\rho,\rho',\tau)\,=\,\frac{1}{N_c^2}
 \int_L \frac{\rmd j}{2 i }\,\frac{1+\rme^{-i\pi j}}{\sin \pi j}\,
 \frac{\rme^{\tau (j-1)}}
 {\sqrt{j-j_0}}\,\exp\left\{(\rho-\rho')\left(\frac{j-2}{2} -
 \sqrt{\frac{\sqrt{\lambda}}{2}\,(j-j_0)}\right)\right\}\,,\nn
 \eeq
where we have used Eqs.~(\ref{gammaP}) and (\ref{nuj}) to explicit the
function $\gamma(j)$. The contour $L$ runs parallel to the imaginary $j$
axis and crosses the real axis between the branch point at $j=j_0\equiv
2-{2}/{\sqrt{\lambda}}$ and the pole at $j=2$ (see Fig. \ref{contour}).

  \begin{figure}
\begin{center}
\includegraphics[width=16cm]{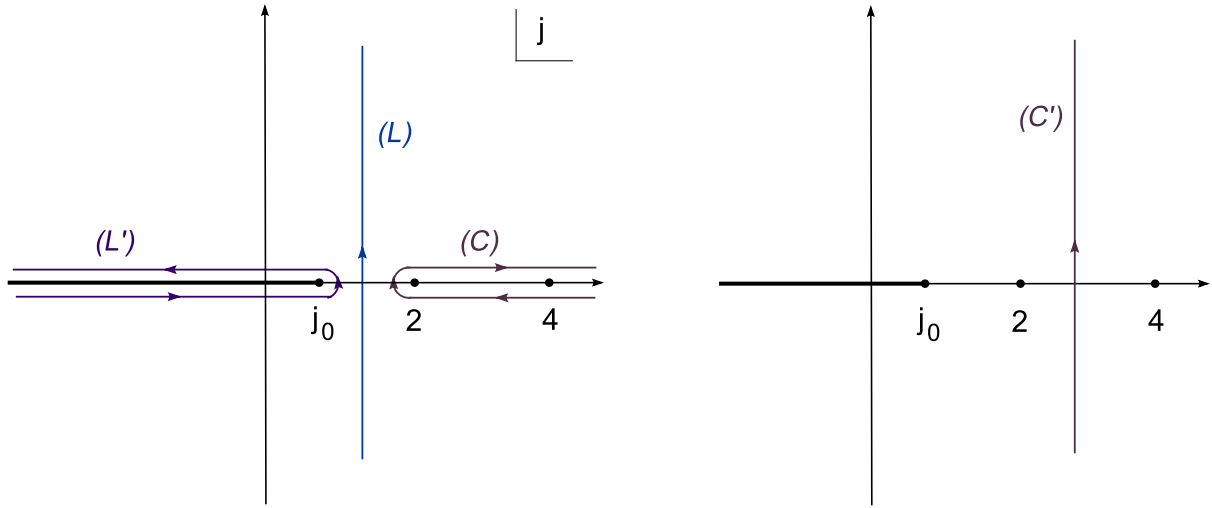}
\caption{\sl The analytic structure of the integrand in Eq.~(4.24)
in the complex $j$ plane, together with various
contours used when evaluating the integral.}
 \label{contour}
\end{center}
\end{figure}

For sufficiently high energies, the integral can be evaluated by
deforming the contour $L\to L'$, with $L'$ encircling the branch cut that
ends at $j_0$, as shown in Fig. \ref{contour}. Alternatively, and more
conveniently for our present purposes, one can use the saddle point
method. Namely, the integrand has a saddle point at
 \beq\label{saddleP}
  j^*\,\simeq\,j_0\,+\,
 \frac{\sqrt{\lambda}(\rho-\rho')^2}{8\tau^2}\,,
 \eeq
where we have used $\tau\gg\rho-\rho'$, as appropriate at high energy. So
long as $\sqrt{\lambda}(\rho-\rho')/4 < \tau$, this saddle point lies
between $j_0$ and 2, and then the use of the saddle point approximation
is indeed justified. By also using $\rme^{-i\pi j_0}\approx 1+2\pi
i/\sqrt{\lambda}$, one thus finds that the high--energy amplitude
develops an imaginary part\footnote{The Virasoro--Shapiro formula in flat
space also gives an imaginary part in a similar manner. $\rme^{-i\pi
\alpha't/2}/t\approx 1/t-i \pi \alpha' /2$. This is how Eq.~(\ref{st})
was actually derived.} $T_\mathbb{P}\equiv{\rm Im}\,\tilde T_\mathbb{P}$,
with
\beq\label{pomeronstring}
 T_\mathbb{P}(\rho,\rho',\tau)\,\approx\,\frac{1}{N_c^2}\,
 {\rme^{\omega_0 \tau}}\
 \rme^{-(1-\gamma_0)(\rho-\rho')}\,\rme^{-\frac{(\rho-\rho')^2}{4D\tau}}
 \,,
 \eeq
where  $\omega_0 = 1-{2}/{\sqrt{\lambda}}$, $\gamma_0
=1-{1}/{\sqrt{\lambda}}$, and $D= {2}/{\sqrt{\lambda}}$.

Eq.~(\ref{pomeronstring}) coincides with the formula previously used in
Sect. 3, cf. Eqs.~(\ref{last}) or (\ref{GGT}), up to a factor
$\rme^{-(1-\gamma_0)\rho}= \left(u'^2/u^2\right)^{1/\sqrt{\lambda}}\,$.
This additional factor has the same origin as the factor
$(r^2/r'^2)^{1/2}$ in the BFKL case, Eq.~(\ref{TPOMERON}) --- namely, is
reflects the high--energy `anomalous dimension' ---, but it is less
important here because the anomalous dimension $\gamma_0$ is close to one
(in contrast to the BFKL case where $\gamma_0=1/2$). And indeed, by
inspection of Eq.~(\ref{pomeronstring}), one can check that the
additional term $(1\!-\!\gamma_0)(\rho-\rho')$ in the exponent remains
negligible so long as $\rho-\rho'\ll \tau$, and hence it cannot affect
our previous conclusions in Sect. 3.2. Note also that, in this
high--energy approximation, the graviton pole $\sim 1/t$ is absent in the
real part of the string amplitude --- in contrast to what happens in flat
space ---, because of the shift in the Pomeron intercept by the
space--time curvature: $\sin \pi j_0 \neq 0$.

However, the graviton pole reappears when moving to sufficiently large
values of $\rho$ (for a given $\tau$), as we explain now. Indeed, when
 \beq \label{region} \frac{\sqrt{\lambda}}{4}(\rho-\rho')
 > \tau \,, \eeq
the saddle point $j^*$ exceeds 2, and then one has to deform the contour
from $C$ to $C'$ (see Fig. \ref{contour}), in order to separately
evaluate the pole of $(1/\sin \pi j)$ at $j=2$. This leads to
  \beq  \label{tree}
  \tilde T_\mathbb{P}(\rho,\rho',\tau)\,\approx\,
  \frac{\rme^\tau}{N_c^2}\,\rme^{-(\rho-\rho')} +
  \frac{1}{N_c^2}\int\limits_{C'}
\frac{\rmd j}{2 i}
 \frac{1+e^{-i\pi j}}{\sin \pi j}
 \frac{\rme^{(j-1)\tau}}{\sqrt{j-j_0}}\,
 \rme^{-(1-\gamma(j))(\rho-\rho')}
 \,. \eeq
The first, purely real, term $\propto {\rme^\tau} (u'^2/u^2)$ is
recognized as the propagator of the elementary, massless, graviton in
$AdS_5$. For large $\tau$ and $\rho$ (such that the condition
(\ref{region}) is satisfied), this pole term dominates over the remaining
contour integral. Indeed, the function $f(j)\equiv (j-1)\tau +
\gamma(j)(\rho-\rho')$
obeys $f(2)=\tau$, $f'(j^*)=0$, and $f'(j) <0$ for any real $j$ within
the interval $2\le j < j^*$. Hence, whenever (\ref{region}) is satisfied,
one can choose the contour $C'$ (see Fig. \ref{contour}) to cross the
real axis at some $j_1$ with $2< j_1 < j^*$, and then ${\rm Re}\,f(j) <
\tau$ for any $j$ along $C'$.

Thus for high energy and sufficiently large $\rho$, the `Pomeron'
exchange is dominated by the elementary graviton and the amplitude is
predominantly real. Since the graviton propagator is very non--local in
$u$, it can connect the large distance between $u\sim QR^2$ and $u'\sim
\Lambda R^2$ without much suppression. Accordingly, with increasing
$\tau$, the corresponding amplitude becomes of $\order{1}$ relatively
fast (much faster than for the diffusive Pomeron in
Eq.~(\ref{pomeronstring}) !), namely when
 \beq \label{line} \tau\,=\, \rho-\rho'+\ln N_c^2\,, \eeq
in agreement with Eq.~(\ref{tausat}). Of course, being real, this single
graviton exchange does not contribute to $F_2$. Yet, multiple such
exchanges can do so, as we shall shortly explain.

To summarize, the single--Pomeron--exchange approximation to $F_2(x,Q^2)$
is expected to be a legitimate approximation at high energy and strong
coupling so long as $\rho \equiv \ln Q^2/\Lambda^2 <
4\tau/\sqrt{\lambda}$ and for rapidities $\tau <\tau_s(\rho)$, with the
saturation rapidity $\tau_s(\rho)$ given by Eq.~(\ref{taus}). The two
delimitating curves intersect with each other at the point
$(\rho_c,\tau_c)$, Eq.~(\ref{inter}). This is also the point where starts
the large--$\rho$ branch of the saturation line, cf. Eqs.~(\ref{tausat})
or (\ref{line}), to be further discussed in the next subsection.

\subsection{Saturation line from multiple graviton exchanges}

The previous analysis shows that the physics of unitarization in DIS at
strong coupling and relatively large $Q^2$ is conceptually different from
what happens either at weak coupling, or at strong coupling but for lower
values of $Q^2$. In the last two cases, which are the more standard ones,
the black disk limit for the scattering amplitude is approached by
unitarizing the single Pomeron exchange, which in turn requires energies
which are high enough for the two scales involved in the collision to be
joint by diffusion. But for strong coupling and large $Q^2$, the single
Pomeron exchange is unable to saturate the energy--momentum sum rule
(\ref{tauint})--- essentially, because its imaginary part decays too fast
at large $Q^2$. However, precisely in that large--$Q^2$ region, the full
Pomeron propagator in string theory develops an additional, real, part,
which describes elementary graviton exchanges and has the potential to
generate stronger interactions, since very non--local in $u$. This new
component can contribute to the DIS structure functions too, but only via
multiple scattering. In this subsection, we would like to argue (without
performing a fully fledged calculation of  multiple scattering in
$AdS_5$) that multiple graviton exchanges dominate indeed at high energy
and large $Q^2$ (in the vicinity of the saturation line) and, in
particular, they saturate the sum rule in the way already suggested after
Eq.~(\ref{Isat}).

The presence of a large real part in the Pomeron propagator describing
the massless graviton exchange is a common feature in string theories,
and hence also in gauge theories with a gravity dual. Multiple scattering
in such theories is described by multi--loop diagrams, and at each order
of this loop expansion the dominant contribution is provided by the
exchange of $n$ gravitons (for a $(n\!-\!1)$--loop amplitude). It is
well--known that in flat space such higher loop terms exponentiate in the
eikonal approximation \cite{hooft,amati}, and the same was recently shown
to hold for high energy scattering in $AdS_5$ \cite{tan,cor}. The eikonal
approximation breaks down at very small impact parameters and/or at
extremely high energies (see below). In a string dual description of DIS
at strong coupling, the impact parameter in the fifth dimension is large
$u/u' \sim Q/\Lambda$, and therefore the eikonal resummation is expected
to work reasonably well.

In general, the exponentiation is done in the three--dimensional impact
parameter space spanned by the radial coordinate $u$ and the
two--dimensional, physical, impact parameter $\vec{b}$ (neglecting
nonlocality on $S^5$), whereas our previous formul\ae{}, like
Eq.~(\ref{SST}), are already integrated over $\vec{b}$. While it is
possible to undo this integral and analyze the three--dimensional
propagator following \cite{tan,cor}, one may anticipate, following the
example of QCD at weak coupling (cf. Sect. 2; see also Ref. \cite{hm}),
that a simplification occurs for relatively central impact parameters,
$b\ll 1/\Lambda$. DIS is characterized by the large resolution scale
$Q^2\gg \Lambda^2$ which determines the size of the `active' region in
the transverse space. Hence, so long as $b\ll 1/\Lambda$ one can neglect
edge effects and work with scattering amplitudes averaged over $b$, so
like Eq.~(\ref{SST}).


In the present context, the eikonal approximation amounts to
exponentiating the tree--level `Pomeron' amplitude Eq.~(\ref{SST}), which
thus becomes the phase shift in the $S$--matrix: $S=\exp\{i\tilde
T_\mathbb{P}\}$. In the interesting domain at large $\rho\equiv
\ln(Q^2/\Lambda^2)$, namely, $\rho > 4\tau/\sqrt{\lambda}$, the Pomeron
amplitude is given by Eq.~(\ref{tree}) and has both real and imaginary
parts.

The {\em imaginary} part describes `inelastic processes' in which the
final states are (generally massive) string excitations, as obtained by
cutting through the Pomeron. This imaginary part can be evaluated in the
saddle point approximation, with the result displayed in
Eq.~(\ref{pomeronstring}). By itself, this contribution would generate
the saturation line in Eqs.~(\ref{rhosGD0}) or (\ref{tausat}), which
however has been argued to be inconsistent at large $\rho>\rho_c$, in
Sect. 3.2.

The relevant contribution at large $\rho$ is rather the one generated by
the {\em real} part of the phase--shift, that is, by the exponentiation
of the elementary graviton exchange. The corresponding imaginary part
starts at the level of the two--graviton exchange, where it describes, in
particular, the cross--section for the elastic scattering between two
states localized at $u$ and $u'$, respectively. Note however that, in the
context of DIS, this process is not truly elastic --- rather, it is a
particularly simple {\em diffractive} process, which contributes to $F_2$
---, because the incoming virtual photon (or ${\mathcal R}$-current)
does not appear in the final state. The amplitude for such a double
graviton exchange is estimated as
 \beq\label{quel} T_{2g}(\rho,\rho',\tau)\,\approx\,\left[
  \frac{\rme^\tau}{N_c^2}\,\rme^{-(\rho-\rho')}\right]^2\,,\eeq
and becomes of $\order{1}$ along the saturation curve (\ref{line}). More
general diffractive processes --- characterized by a rapidity gap between
the products of the collision --- are obtained by `cutting' the multiple
scattering series in between successive graviton exchanges, and by
allowing for more general diffractive states: the strings in the final
states need not be exactly the same as the incoming ones, rather they can
be internal excitations of the latter, like higher Kaluza--Klein modes.
Whenever this happens, the process is genuinely absorptive, already at
the level of the elementary $\rho-\rho'$ amplitude.

The diffractive excitations of the string have been first discussed in
Ref.~\cite{amati} in flat space. Leaving a rigorous treatment of this
effect in $AdS_5$ for future work, here we only give an estimate of its
magnitude. The internal excitation of incoming strings is due to the
finiteness of the string size. This effect can be understood as the
result of the tidal force \cite{gid} experienced by one of the strings
traversing a gravitational shock wave created by the other. In the
operator eikonal approach of \cite{amati}, the real part of the phase
shift reads
  \beq  \frac{\rme^\tau}{N_c^2}\frac{U^{'2}}{U^2} \approx
   \frac{\rme^\tau}{N_c^2}\frac{u'^2}{u^2} +
   \frac{1}{2}\frac{\partial^2}{\partial u^2}
   \left( \frac{\rme^\tau}{N_c^2}\frac{u'^2}{u^2} \right)
   \tilde{U}\tilde{U} +\cdots\,, \label{ex} \eeq
where again capital letters denote operators, decomposed between zero and
non--zero modes as $U=u+\tilde{U}$. The non--zero mode $\tilde{U}$ is
typically of the order the string size, as measured by a local inertial
observer:
 \beq \tilde{U} \sim l_s \frac{u}{R}\,, \eeq
so the expansion parameter in Eq.~(\ref{ex}) is
$l_s^2/R^2=1/\sqrt{\lambda}$. The strength of string excitations leading
to absorption is measured by the second term of Eq.~(\ref{ex}). It
reaches unity when
 \beq
 \tau=\rho-\rho'+\ln N_c^2 + \ln\sqrt{\lambda}\,, \eeq
which, once again, is parametrically the same as Eq.~(\ref{line}).
(Remember that we are in the regime where $\ln N_c^2 > \lambda\gg \ln
\lambda$.) We thus see that all types of diffractive processes --- both
the quasi--elastic ones, cf. Eq.~(\ref{quel}), and the genuinely
absorptive ones, cf. Eq.~(\ref{ex}) --- contribute on equal grounds to
the imaginary part of the forward scattering amplitude, and approach the
unitarity limit along the curve exhibited in Eq.~(\ref{line}). This is in
contrast to the flat space result \cite{amati} where the imaginary part
due to string excitations is suppressed with respect to the real part by
an inverse power of the impact parameter $l_s^2/b^2$.


So far, the diffractive unitarity line (\ref{line}) has been derived for
the scattering between two states localized at $\rho$ and $\rho'$. But it
is easy to check that a similar line, with $\rho'=0$, applies to the
dilaton as whole. Indeed, after performing the convolution with the
dilaton wavefunction, the relevant amplitude can be evaluated as in
Eq.~(\ref{TGD0}), which gives
 \beq\label{TGD3}
  T_{2g}(\rho,\tau)
  \,\simeq\,\int\limits_0^{\rho-\rho_{s0}}\rmd\rho'
  \,\left[
  \frac{\rme^\tau}{N_c^2}\,\rme^{-(\rho-\rho')}\right]^2
   \,{\rm e}^{-\Delta\rho'}
 \,+\, \int\limits_{\rho-\rho_{s0}}^\infty
 \rmd\rho'  \,{\rm e}^{-\Delta\rho'}\,,
 \eeq
where now $\rho_{s0}\equiv \tau - \ln N_c^2$, cf. Eq.~(\ref{line}), and
we have assumed that $\rho > \rho_{s0}$. In writing the equation above,
we have focused on the two--graviton exchange, cf. Eq.~(\ref{quel}),
since this is the dominant contribution when approaching the saturation
lien from large values of $\rho$. At this point, it is important to
remember that the dilaton conformal dimension satisfies $\Delta > 2$, as
manifest on Eq.~(\ref{Delta}). Hence, the first integral in
Eq.~(\ref{TGD3}) is dominated by its lower limit $\rho'=0$ and, moreover,
the second integral is exponentially suppressed w.r.t. the first one.
Thus, once again, we find that the scattering is controlled by the
dilaton component living near the infrared cutoff at $u\sim u_0$. This
yields
 \beq\label{TGD4}
  T_{2g}(\rho,\tau)
  \,\simeq\,\left[
  \frac{\rme^\tau}{N_c^2}\,\rme^{-\rho}\right]^2\,,\eeq
which becomes of $\order{1}$ for $\rho\simeq \rho_s(\tau)= \tau - \ln
N_c^2$, in agreement with Eq.~(\ref{tausat}).

For $\rho > \rho_c$, with $\rho_c$ defined in Eq.~(\ref{inter}), the
curve (\ref{tausat}) lies below the unitarity line (\ref{taus}) for
inelastic processes (see also Fig. \ref{phase2}), meaning that, when
increasing $\tau$ in this regime, the unitarization of DIS proceeds via
diffractive processes. Therefore, Eq.~(\ref{tausat}) represents the {\em
physical saturation line} for $\rho > \rho_c$. This conclusion is further
corroborated by the following facts: \texttt{(i)} at $\rho = \rho_c$, the
two branches of the saturation line, Eq.~(\ref{taus}) and, respectively,
Eq.~(\ref{tausat}), continuously match with each other, and \texttt{(ii)}
for $\rho > \rho_c$, the energy--momentum sum rule (\ref{tauint}) is
saturated by the diffractive processes (see below), showing that no
important physical mechanism has been left out by the previous analysis.

To evaluate the sum--rule (\ref{tauint}), notice first that a
contribution to $F_2$ due to the exchange of $n$ gravitons (with $n\ge
2$), is parametrically
 \beq\label{like}
 F_2(\tau,Q^2)\ \sim \ N_c^2\,\rme^{\rho}\,T_{ng}(\rho,\tau)
 \ \sim \ N_c^2\,\frac{Q^2}{\Lambda^2} \,
 \left(\frac{\rme^{\tau}}{N_c^2(Q^2/\Lambda^2)}\right)^n\,. \eeq
When this is inserted within the Eq.~(\ref{tauint}), it is clear that,
for any $n\ge 2$, the integrand is strongly peaked at the saturation
value $\tau_s=\rho+\ln N_c^2$: indeed, the integrand rises exponentially
with $\tau$ so long as $\tau <\tau_s$, but it is exponentially decreasing
at $\tau >\tau_s$. Moreover, at the peak value,
$T_{ng}(\rho,\tau=\tau_s)\sim\order{1}$, so the integral is evaluated as
 \beq\label{sumhigh} \int \rmd \tau\, \rme^{-\tau}
 F_2(\tau,Q^2)\ \sim \ N_c^2\,\frac{Q^2}{\Lambda^2}\ \rme^{-\tau_s(\rho)}\
 \sim\,\order{1}.\eeq
This result is independent of $Q^2$, as anticipated. Interestingly, the
above calculation also shows that, at high $Q^2$, most of the total
energy (or longitudinal momentum) of the hadron lies along the saturation
line (\ref{tausat}), that is, at relatively large rapidities, or small
values of $x$. This is in sharp contrast with what happens in QCD at weak
coupling, where the energy of a fast moving hadron is predominantly
carried by the hadron constituents (`partons') with relatively large
values of $x$, or small values of $\tau$, well below the corresponding
saturation value $\tau_s(\rho)$. We shall return to this observation in
the final, discussion, section, where we shall attempt a partonic
interpretation for our results at strong coupling.

We conclude this section with some estimates for $F_2(\tau,Q^2)$ in the
vicinity of the saturation line (\ref{tausat}), to be compared to the
corresponding results in pQCD, cf. Eqs.~(\ref{f2}) and (\ref{f2sat}).
Note first that Eq.~(\ref{tausat}) implies $Q_s^2(\tau)= \Lambda^2
({\rme^\tau}/{N_c^2})$.

For relatively low virtualities, $Q^2\simle Q_s^2(\tau)$, one finds
 \beq \label{F2low}  F_2(\tau,Q^2)
 \sim \frac{N_c^2Q^2}{\Lambda^2}\qquad\mbox{when}\qquad
 Q^2\simle Q_s^2(\tau)\,, \eeq
which shows \emph{saturation}, in the sense that $F_2$ does not increases
with $\tau$, and that $F_2/Q^2$ is not divergent as $Q^2\to 0$. (A
physical interpretation for this behaviour will be proposed in Sect. 5.)

For $Q^2$ much larger than $Q_s^2(\tau)$, the structure function is
dominated by the two--graviton exchange and scales as
 \beq\label{we0}
 F_2(\tau,Q^2)\sim \frac{N_c^2Q^2}{\Lambda^2}\left(\frac{\rme^\tau
 \Lambda^2}{N_c^2 Q^2}\right)^2\,, \eeq or
 \beq
 \frac{F_2(\tau,Q^2)}{Q^2}\sim
 \frac{N_c^2}{\Lambda^2}\left(\frac{Q_s^2(\tau)}{Q^2} \right)^2
 \qquad\mbox{when}\qquad
 Q^2\gg Q_s^2(\tau)\,.
 \label{we}  \eeq
This is \emph{geometric scaling} (compare to the weak--coupling result in
Eq.~(\ref{f2})), with an effective anomalous dimension $\gamma_s=-1$,
which however is no longer associated with the anomalous dimension of the
twist--two operator. Indeed, the contribution exhibited in
Eqs.~(\ref{we0}) or (\ref{we}) is obviously of higher twist order, and in
fact this process has an analog in standard DIS: the `diffractive
dissociation' of the virtual photon, $\gamma^*p\to q\bar qp$, which
involves the elastic scattering between the `color dipole' (the $q\bar q$
pair) and the proton (see, e.g., \cite{FR97,GBW99,HIMST06}). There are
however some interesting differences between these results and the
corresponding ones at weak coupling:

 \texttt{(i)} In pQCD, the
diffractive scattering represents only a tiny fraction of the total DIS
cross--section, whereas in the present, strong--coupling, case this has
been found to be the dominant component at high energy, which in
particular is responsible for unitarization.

\texttt{(ii)} The diffractive structure function $F_2^D$ in pQCD at high
energy (small $x$) and large $Q^2$ is formally a leading--twist quantity,
$F_2^D/Q^2\sim 1/Q^2$, although the relevant dipole amplitude is, of
course, of higher--twist order. This is so since, in computing $F_2^D$,
the convolution with the virtual photon wavefunction, Eq.~(\ref{PsiTL}),
is dominated by the relatively large $q\bar q$ excitations having $r\gg
1/Q$ (the so--called `aligned--jet configurations') \cite{GBW99,HIMST06}.
This does not happen, however, in the present context at strong coupling,
where the convolution in Eq.~(\ref{factori}) is controlled by $u\sim
u_c\sim Q$. Accordingly, the ensuing structure function is a
higher--twist quantity, cf. Eq.~(\ref{we}), so like the elementary
scattering amplitude in Eq.~(\ref{quel}).

\section{Physical discussion: Towards a partonic picture at strong coupling}
\setcounter{equation}{0}

Let us first summarize the main new results in this paper, before we
propose a physical interpretation for them. (These results are also
graphically summarized in Fig. \ref{phase2}.) By studying deep inelastic
scattering at strong 't Hooft coupling and high energy within the dual
string theory, we have inferred the position of the saturation line which
separates the weak--scattering regime from the strong--scattering one in
the kinematical plane $\tau-\rho$, with $\tau=\ln(s/Q^2)$ and
$\rho=\ln(Q^2/\Lambda^2)$. This line is in fact the juxtaposition of two
curves, matching with each other at the point $(\rho_c,\tau_c)$ defined
in Eq.~(\ref{inter}), which correspond to the two different mechanisms
responsible for unitarization in different kinematical regions :

 \begin{figure}[h]
\begin{center}
\includegraphics[width=13.cm,bb=100 230 650 750]{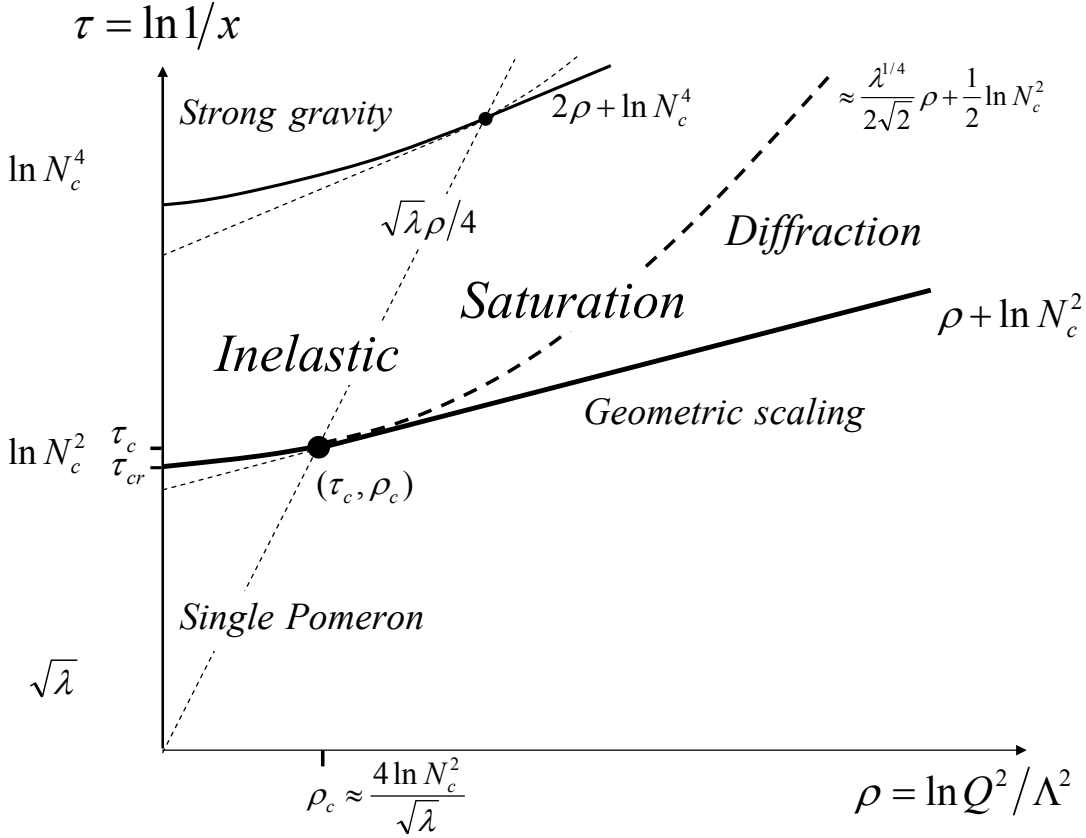}
\caption{\sl Proposed `phase diagram' for DIS at high energy and
strong coupling when $\,\ln N_c^2 \gg \sqrt{\lambda}$.
The continuous, thick, line represents the saturation line.
The other boundaries are explained in the text.
 \label{phase2}}
\end{center}
\end{figure}

\vspace*{0.2cm}
\begin{itemize}
\item For $\rho\le \rho_c\simeq 4\ln N_c^2/\sqrt{\lambda}$, the
    scattering amplitude in the vicinity of the unitarity limit is
    controlled by a single Pomeron exchange; this amplitude becomes
    of $\order{1}$ on the saturation line given by
    Eqs.~(\ref{rhosGD0}) or (\ref{taus}), i.e., when $Q^2=
    Q_s^2(\tau)$, with the {\em saturation momentum}
   \beq
  Q_s^2(\tau)\,=\,{\Lambda^2}\, \rme^{\sqrt{4D\tau(\omega_0 \tau-\ln
  N_c^2)}}\qquad \mbox{for}\qquad\tau_{\rm cr} < \tau<\tau_c\,,
  \label{Qslow}
  \eeq
with  $\omega_0 = 1-{2}/{\sqrt{\lambda}}$, $D= {2}/{\sqrt{\lambda}}$,
and $\tau_{\rm cr} \simeq \ln N_c^2$ is the critical rapidity for the
onset of unitarity corrections.

\vspace*{0.2cm}\item For $\rho > \rho_c$, the dominant contribution
    to the DIS structure function before unitarization comes from
    multiple exchanges of elementary gravitons; the reaches the
    unitarity limit on the saturation line shown in Eq.~(\ref{taus}),
    i.e., for $Q^2= Q_s^2(\tau)$, with
    \beq
  Q_s^2(\tau)\,=\,\Lambda^2\, \frac{\rme^\tau}{N_c^2} \qquad
  \mbox{for}\qquad\tau >\tau_c\,. \label{Qshigh}
  \eeq

\end{itemize}

\noindent In Fig.~\ref{phase2}, the lower branch of the saturation line,
Eq.~(\ref{Qslow}), is also represented, as a dotted line, for
$\tau>\tau_c$, where it indicates the `unitarity limit' for inelastic
processes, i.e., the line along which the contribution of the
single--Pomeron exchange to the DIS amplitude becomes of $\order{1}$.
Note that the `diffusion approximation' that was implicitly used in
deriving the equation of this line in Sect. 3.2 is still
valid\footnote{As discussed in Sect. 4.1, the diffusion approximation is
expected to be correct so long as $j\ll\sqrt{\lambda}$.} for such large
values of $\rho$ because,  along this line, $j\simle 3$, with the upper
limit $j=3$ corresponding to $\rho\to\infty$. (This can be checked by
using Eq.~(\ref{saddleP}) together with the large--$\rho$ version of
Eq.~(\ref{taus}), that is, $\tau_s(\rho)\simeq
(\lambda^{1/4}\rho/\sqrt{2} + \ln N_c^2)/2$.) Below that line and for
$\rho > \rho_c$, the DIS cross--section is dominated by the diffractive
processes (whose contribution becomes of $\order{1}$ already along the
proper saturation curve, Eq.~(\ref{Qshigh})), whereas above that line,
both diffractive and inelastic processes contribute on equal footing and
together give an amplitude $T=1$. Also, below the saturation line
(\ref{Qshigh}), the DIS cross--section for $\rho > \rho_c$ is
predominantly diffractive, of higher--twist order, and it exhibits
geometric scaling, cf. Eq.~(\ref{we}). The upper corner of Fig.
\ref{phase2} denoted as `strong gravity' will be discussed towards the
end of this section.

So far we have assumed $N_c^2 > \rme^{\sqrt{\lambda}}$, to allow both
regimes alluded to above to coexist with each other. But it should be
clear that the opposite case $N_c^2 \ll \rme^{\sqrt{\lambda}}$ (with $N_c
\gg 1$, though) is included too in our results, as the special limit
$\rho_c\to 0$\,: then, the first regime squeezes to zero, meaning that
there is not enough phase--space for the genuine Pomeron behaviour to
appear --- the amplitude reaches values of order one via multiple
graviton exchanges already before the diffusion becomes important.

Note that the above results do not involve special properties of the
dilaton target, like its mass or its conformal dimension $\Delta$. We
therefore expect them to universally apply (at strong coupling) to all
hadrons which, like the dilaton, have a dual state localized near the
infrared cutoff $u_0=\Lambda R^2$ in the radial dimension of $AdS_5$. The
same universality applies to the form of the structure function
$F_2(\tau,Q^2)$ in the vicinity of the saturation line. In the
weak--scattering regime at $Q^2\gg Q_s^2(\tau)$, $F_2$ shows a rapid rise
with $\tau$ and a rapid decrease with $Q^2$, $F_2\propto
(\rme^{2\tau}/Q^2)$, cf. Eq.~(\ref{we0}), which is the hallmark of the
double--graviton exchange. Furthermore, at low momenta $Q^2\simle
Q_s^2(\tau)$, one finds $F_2\sim {N_c^2Q^2}/{\Lambda^2}$, cf.
Eq.~(\ref{F2low}), a structure which is an almost automatic consequence
of the definition (\ref{F2u}) of $F_2$ together with the black disk limit
for the scattering amplitude. But in QCD at weak coupling at least, this
structure is also tantamount of {\em parton saturation}, which makes it
tempting to propose a similar interpretation in the strong--coupling case
as well.

Let us first briefly recall the situation in pQCD (see, e.g., the review
papers \cite{SATreviews} for more details): there, the structure function
$F_2(\tau,Q^2)$ measured in DIS can be interpreted as the {\em quark
distribution} in the target, as measured in the special frame (the
`infinite momentum frame', or IMF) in which the target has a very large
longitudinal momentum $P\gg M$. The `quark distribution' is the number of
quarks per unit rapidity which are localized in impact parameter space
within an area $\sim 1/Q^2$ fixed by the resolution of the virtual
photon; alternatively
--- via the uncertainty principle ---, this is the number of quark
excitations with transverse momenta $k_\perp \simle Q$. Hence, by knowing
$F_2(\tau,Q^2)$, one can estimate the {\em quark occupation number} as
  \beq\label{nq} n_q(\tau,k_\perp)
  \,\equiv\,\frac{\rmd N_q}{\rmd\tau \,\rmd^2 k_\perp \rmd^2 b_\perp}
 \,\simeq\, \frac{1}{N_c}\, \frac{\rmd F_2}{\rmd Q^2\,\pi R_0^2}
 \,,
 \eeq
where the derivative in the r.h.s. is evaluated at $ Q^2=k_\perp^2$,
$R_0$ is the radius of the hadron disk, and the factor $1/N_c$ enters
since we consider quarks of a given color. (Other numerical factors, like
the number of spin states, or the number of flavors, are not explicitly
shown.) Then, by using the estimates (\ref{f2}) and (\ref{f2sat}) for
$F_2$, one can distinguish between two physical regimes: \texttt{(i)} a
dilute regime at high momenta $k_\perp \gg Q_s(\tau)$, where
$n_q(\tau,k_\perp)$ scales like\footnote{For very large $k_\perp^2$, the
anomalous dimension dies away, $\gamma_s\to 0$, and one recovers the
expected bremsstrahlung spectrum $n_q\propto 1/k_\perp^2$.} $n_q\sim
({Q_s^2(\tau)}/{k_\perp^2})^{1-\gamma_s}$, and thus rises rapidly with
$\tau$ due the BFKL evolution, and \texttt{(ii)} a saturation regime at
momenta $k_\perp \simle Q_s(\tau)$, where $n_q$ saturates at a value of
order 1. This is {\em quark saturation} and, as explained in Sect. 2, it
ultimately reflects the saturation of the {\em gluon} occupation numbers
in pQCD at small $x$ \cite{GLR,MQ85,MV,AM99,SAT} : $n_g(\tau,k_\perp)\sim
1/\abar$ for $k_\perp \simle Q_s(\tau)$.

Returning to the strong--coupling problem of interest, it is clear that a
straightforward partonic interpretation is now hindered by the lack of a
simple relation between the operators which enter the structure of the
$\mathcal{R}$--current (for the ${\mathcal N}=4$ SYM theory, this current
receives contributions from the Weyl fermions and from the scalar fields)
and the physical excitations in the target wavefunction: the latter would
be created, or annihilated, by the respective fundamental fields only to
lowest order in perturbation theory; but at strong coupling this relation
could be strongly renormalized, in a way which is not under control. It
is not clear, though, whether such a renormalization will completely wash
out the parton picture suggested by perturbation theory. For instance, it
is known that the entropy of strongly coupled ${\mathcal N}=4$ SYM theory
at finite temperature, which is a direct measure of the number of states,
differs only by a factor 3/4 from the entropy of the corresponding ideal
gas, where all the states are associated with free fundamental fields.
This is suggestive that, at least, the counting of sates is quite similar
at strong coupling and at weak coupling.

A similar suggestion emerges from our present results for DIS at strong
coupling. When analyzed in view of Eq.~(\ref{nq}) (with $N_c\to
N_c^2-1\approx N_c^2$, to account for the proper number of color degrees
of freedom of the fields in ${\mathcal N}=4$ SYM), the estimate
(\ref{F2low}) for $F_2$ suggests that, for transverse momenta $k_\perp
\simle Q_s(\tau)$, the occupation numbers for the fermionic and scalar
fields saturate at a value of order 1. Note that, for fermions, this
value $n=1$ is in agreement with the maximal occupancy allowed by the
Pauli principle. Our analysis gives no direct access to the gluon
occupancy, but in view of the supersymmetry of the underlying theory, we
expect a similar saturation value, $n_g\sim 1$ for $k_\perp \simle
Q_s(\tau)$, for the gluons as well. This value is quite reasonable: at
small coupling, one needs a high occupancy $n_g\sim 1/\abar \gg 1$ in
order to compensate for the weakness of the interactions, but when the
coupling is strong, the gluons mutual repulsion responsible for
saturation becomes important already when $n_g\sim 1$.

By pushing this interpretation towards larger virtualities $Q^2\gg
Q_s^2(\tau)$, where Eq.~(\ref{we0}) applies, one finds that, when
increasing $k_\perp$ above $Q_s(\tau)$, the (scalar and fermion)
occupation numbers decrease very fast, as $n\sim
({Q_s^2(\tau)}/{k_\perp^2})^{2}$. Assuming that this behaviour extends to
gluons as well, we deduce that, at strong coupling, there are essentially
no partons at transverse momenta above the saturation scale: at large
$k_\perp$, the spectrum converges so fast that any interesting
convolution involving the integral of $n(\tau, k_\perp)$ over $k_\perp$
will be dominated by $k_\perp \simle Q_s(\tau)$. Equivalently, when the
hadron is probed with a given transverse resolution scale $Q^2$, almost
all partons appear to live at very small values of $x$, or large values
of $\tau\equiv \ln(1/x)$, above the saturation line: $\tau
>\tau_s(Q^2)$. Since, on the other hand, such small--$x$ partons
carry only little longitudinal momentum, it is natural to find that the
energy--momentum sum rule is dominated by those partons living along the
saturation line, as shown by Eq.~(\ref{sumhigh}).

\comment{ At this point it is appropriate to make contact with some
previous approaches in the literature, especially, the closely related
one by Polchinski and Strassler \cite{dis}. As already mentioned in Sect.
3, the situation considered in Ref. \cite{dis} corresponds to the strict
large--$N_c$ limit, where there is no unitarity issue --- the scattering
amplitude is always smaller than one --- and the structure function at
small $x$ (with $\ln(1/x) \simge \sqrt{\lambda}$) is correctly given by
the single--Pomeron--exchange approximation, which provides an universal
result at high energy (cf. Eq.~(\ref{TSP})). For not so small values of
$x$, on the other hand, DIS is controlled by protected operators of the
higher--twist type, and $F_2$ is not universal anymore: at large $Q^2$,
it decays like $F_2\sim (1/Q^2)^{\Delta-1}$, with $\Delta$ the conformal
dimension of the dilaton. Note that a similar decay at large $Q^2$ shows
up already in regime controlled by the single--Pomeron exchange, when
this is considered for not so small values of $x$ and relatively large
values of $\rho\sim \ln Q^2$, cf. Eq.~(\ref{TGD2}). This suggests that
there might be possible to perform a smooth matching between our present
results at small $x$ and those derived in Ref. \cite{dis} for larger
values of $x$, via a more detailed study of the transition region. In
particular, such an analysis should determine the boundary of the
geometric scaling region at large $Q^2$ (cf. Fig.~\ref{phase2}), which
remains unknown in the present analysis.}

It is interesting to notice that the `phase diagram' in
Fig.~\ref{phase2}, or, more precisely, a slice of it at fixed $\tau$,
bears some resemblance to a previous scenario for high--energy string
scattering in flat space by Amati {\it et al} \cite{amati}, after
replacing the impact parameter $b$ of Ref. \cite{amati} by the photon
virtuality $Q^2$. Specifically, both pictures have in common the
dominance of the diffractive over the inelastic cross--section at large
values of $\rho$ (with $\rho=\ln Q^2$ in the present context and $\rho=b$
in the case of Ref. \cite{amati}) and for sufficiently large values of
$\tau$. But a closer inspection reveals also some important differences
between these two pictures, which can be attributed to our use of a
nontrivial background metric. In particular, and unlike in Ref.
\cite{amati}, in our present analysis there is no hierarchy between the
elastic and diffractive processes: they all contribute on the same
footing and simultaneously become large on the `elastic line' in
Eq.~(\ref{line}).

Consider finally the corner of the `phase diagram' at ultrahigh energies
$s\gtrsim N_c^4$ which in Fig.~\ref{phase2} is referred to as `strong
gravity'. This is the region where the square of the momentum transferred
in the $u$ direction, $t=k_u k^u = G^{uu}k_uk_u$, which grows with $s$
like $t_u(s)\propto s^2$ (see the Appendix) becomes of the same order as
the total energy squared $s$, meaning that the eikonal approximation
cannot be trusted anymore. Physically, this is generally interpreted (so
far, mostly in the context of flat space analyses
\cite{amati,Veneziano04,Giddings07}) as the onset of non--linear gravity
effects, possibly associated with the formation of a black hole. In this
regime, the graviton self--interactions cannot be neglected anymore, so
in the context of the scattering problem one has to resum an infinite
number of interacting gravitons in the $t$--channel. This is reminiscent
of the non--linear, strong gluon field problem of pQCD in the saturation
regime \cite{MV,JKLW,CGC,SAT,PLOOP,eff}. It would be therefore
interesting to see if the similarity between the two phenomena could be
made even sharper.

\section*{Acknowledgments}
We are grateful to R.~Brower and C.~-I.~Tan for answering our questions
and providing us with their preliminary results on the eikonal
resummation in $AdS_5$. We thank I.~Bena, L.~McLerran, and  R.~Peschanski
for discussions. Y.~H. thanks K.~Hosomichi, R.~Janik, and L.~Lipatov for
helpful conversations. The work of A.~M. is supported in part by the US
Department of Energy. The work of E.~I. in supported in part by Agence
Nationale de la Recherche via the programme ANR-06-BLAN-0285-01.

\appendix
\section{On the breakdown of the eikonal approximation at ultrahigh
energies}

In this Appendix, we discuss the breakdown of the eikonal approximation
due to large momentum transfer at ultrahigh energies. The momentum $k_u$
transferred in the $u$ direction can be estimated as the $u$--derivative
of the real part of the phase--shift $\delta$ in the $S$--matrix
($S=\rme^{i\delta}$): $k_u \sim \partial_u \delta$.

Consider first the regime at relatively large $Q^2$, such that
$\rho\equiv \ln(Q^2/\Lambda^2) > 4\tau/\sqrt{\lambda}$. Then, as
explained in Sects. 4.2 and 4.3, the phase--shift is controlled by the
elementary graviton exchange, which yields
  \beq k_u \sim \partial_u
  \left(\frac{\rme^\tau}{N_c^2}\frac{u_0^2}{u^2}\right)\sim
  \frac{\rme^\tau}{N_c^2}\frac{u_0^2}{u^3}\,,   \eeq
and hence the invariant squared momentum transfer reads
 \beq t\equiv G^{uu}k_uk_u \sim \frac{1}{R^2}\,\left[
  \frac{\rme^\tau}{N_c^2}\,\frac{u_0^2}{u^2}\right]^2
  \,,   \eeq
where we have used $G^{uu}=u^2/R^2$. When this becomes on the order of
the total squared energy $\tilde s = ({R^2}/{u^2})s$ (the rescaling
factor ${R^2}/{u^2}$ is necessary to relate the four dimensional energy
to the ten dimensional one \cite{dis})
 \beq t\,\sim\,
 \frac{R^2}{u^2}s\,, \eeq
the eikonal approximation breaks down. In the context of DIS, this
condition must be considered for $u\sim u_c = QR^2$, cf. Eq.~(\ref{uc}),
which is the value that controls the respective cross--section. This
yields the following critical line
 \beq \tau=2\rho+\ln N_c^4\,,
 \label{black} \eeq
above which one expects strong gravity effects, possibly leading to the
formation of a black hole.

Consider similarly the low--$Q^2$ regime, at $\rho<
4\tau/\sqrt{\lambda}$. Then, the phase shift is associated with the
single Pomeron exchange, and can be estimated as in
Eq.~(\ref{pomeronstring}) (for both the real and the imaginary part):
\beq \delta\sim \frac{1}{N_c^2}\,
\rme^{\omega_0\tau-\frac{\rho^2}{4D\tau}}
\eeq
The $u$--derivative introduces a factor of order $1/u$ (since $\partial_u
\rho = 2/u$), so up to subleading prefactors, one finds
\beq t\equiv G^{uu}(\partial_u \delta)^2 \sim \frac{1}{R^2N_c^4}\,
 \rme^{2\omega_0\tau-\frac{\rho^2}{2D\tau}}\,,\eeq
to be compared with $\tilde{s}\equiv ({R^2}/{u^2_c})s={\rme^\tau}/{R^2}$.
This comparison determines a curve
 \beq
\left(1-\frac{4}{\sqrt{\lambda}}\right)\tau
 -\frac{\sqrt{\lambda}\rho^2}{4\tau} = \ln N_c^4\,, \eeq
which smoothly matches the previous line $\tau=2\rho+\ln N_c^4$, cf.
Eq.~(\ref{black}), at a point
 \beq\label{black1} (\tau,\rho)=\left(\frac{\ln N_c^4}{1-8/\sqrt{\lambda}},
 \frac{4\ln N_c^4}{\sqrt{\lambda}-8} \right)\,. \eeq
Together, the two lines in Eqs.~(\ref{black}) and (\ref{black1})
determines the boundary of the applicability of the eikonal approximation
at ultrahigh energies $s\gtrsim N_c^4$, as illustrated in
Fig.~\ref{phase2}.

\providecommand{\href}[2]{#2}
\begingroup\raggedright

\endgroup

\end{document}